\documentclass[twoside,twocolumn,9pt]{article}
\usepackage{extsizes}
\usepackage[super,sort&compress,comma]{natbib} 
\usepackage[version=3]{mhchem}
\usepackage[left=1.5cm, right=1.5cm, top=1.785cm, bottom=2.0cm]{geometry}
\usepackage{balance}
\usepackage{float}
\usepackage{mathptmx}
\usepackage{sectsty}
\usepackage{graphicx} 
\usepackage{amsmath}
\usepackage{amssymb} 
\usepackage{lastpage}
\usepackage[format=plain,justification=justified,singlelinecheck=false,font={stretch=1.125,small,sf},labelfont=bf,labelsep=space]{caption}
\usepackage{float}
\usepackage{fancyhdr}
\usepackage{fnpos}
\usepackage[english]{babel}
\addto{\captionsenglish}{%
  
}
\usepackage{array}
\usepackage{droidsans}
\usepackage{charter}
\usepackage[T1]{fontenc}
\usepackage[usenames,dvipsnames]{xcolor}
\usepackage{setspace}
\usepackage[compact]{titlesec}
\usepackage{hyperref}
\usepackage{lipsum, babel}
\usepackage[normalem]{ulem}

\usepackage{epstopdf}

\definecolor{cream}{RGB}{222,217,201}

\begin{document}

\pagestyle{fancy}
\thispagestyle{plain}
\fancypagestyle{plain}{
\renewcommand{\headrulewidth}{0pt}
}

\makeFNbottom
\makeatletter
\renewcommand\LARGE{\@setfontsize\LARGE{15pt}{17}}
\renewcommand\Large{\@setfontsize\Large{12pt}{14}}
\renewcommand\large{\@setfontsize\large{10pt}{12}}
\renewcommand\footnotesize{\@setfontsize\footnotesize{7pt}{10}}
\makeatother

\renewcommand{\thefootnote}{\fnsymbol{footnote}}
\renewcommand\footnoterule{\vspace*{1pt}%
\color{cream}\hrule width 3.5in height 0.4pt \color{black}\vspace*{5pt}} 
\setcounter{secnumdepth}{5}

\makeatletter 
\renewcommand\@biblabel[1]{#1}            
\renewcommand\@makefntext[1]%
{\noindent\makebox[0pt][r]{\@thefnmark\,}#1}
\makeatother 
\renewcommand{\figurename}{\small{Fig.}~}
\sectionfont{\sffamily\Large}
\subsectionfont{\normalsize}
\subsubsectionfont{\bf}
\setstretch{1.125} 
\setlength{\skip\footins}{0.8cm}
\setlength{\footnotesep}{0.25cm}
\setlength{\jot}{10pt}
\titlespacing*{\section}{0pt}{4pt}{4pt}
\titlespacing*{\subsection}{0pt}{15pt}{1pt}

\fancyfoot{}
\fancyfoot[RO]{\footnotesize{\sffamily{1--\pageref{LastPage} ~\textbar  \hspace{2pt}\thepage}}}
\fancyfoot[LE]{\footnotesize{\sffamily{\thepage~\textbar\hspace{4.65cm} 1--\pageref{LastPage}}}}
\fancyhead{}
\renewcommand{\headrulewidth}{0pt} 
\renewcommand{\footrulewidth}{0pt}
\setlength{\arrayrulewidth}{1pt}
\setlength{\columnsep}{6.5mm}
\setlength\bibsep{1pt}

\makeatletter 
\newlength{\figrulesep} 
\setlength{\figrulesep}{0.5\textfloatsep} 

\newcommand{\topfigrule}{\vspace*{-1pt}%
\noindent{\color{cream}\rule[-\figrulesep]{\columnwidth}{1.5pt}} }

\newcommand{\botfigrule}{\vspace*{-2pt}%
\noindent{\color{cream}\rule[\figrulesep]{\columnwidth}{1.5pt}} }

\newcommand{\dblfigrule}{\vspace*{-1pt}%
\noindent{\color{cream}\rule[-\figrulesep]{\textwidth}{1.5pt}} }

\makeatother

\twocolumn[
  \begin{@twocolumnfalse}
\vspace{1em}
\sffamily
\begin{tabular}{m{0.5cm} p{18cm} }

$~$& \noindent\LARGE{\textbf{Electronic excitation spectra of cerium oxides: from ab initio dielectric response functions to Monte Carlo charge transport simulations}} \\
\vspace{0.3cm} & \vspace{0.3cm} \\

 & \noindent\large{Andrea Pedrielli\textit{$^{a,b}$}, Pablo de Vera\textit{$^{a}$}, Paolo E. Trevisanutto\textit{$^{c}$}, Nicola M. Pugno\textit{$^{b,d}$}, Rafael Garcia-Molina{$^{e}$}, Isabel Abril{$^{f}$}, Simone Taioli$^{\ast}$\textit{$^{a,g}$}, and Maurizio Dapor$^{\ast}$\textit{$^{a}$}}\\ \\

$~$& \noindent\normalsize{Nanomaterials made of the cerium oxides CeO$_2$ and Ce$_2$O$_3$ have a broad range of applications, from  catalysts in automotive, industrial or energy operations to promising materials to enhance hadrontherapy effectiveness in oncological treatments. 
To elucidate the physico-chemical mechanisms involved in these processes, it is of paramount importance to know the electronic excitation spectra of these oxides, which are obtained here through high-accuracy linear-response time-dependent density functional theory calculations. In particular, the macroscopic dielectric response functions $\bar\epsilon$ of both bulk CeO$_2$ and Ce$_2$O$_3$ are derived, which compare remarkably well with the available experimental data. These results stress the importance of appropriately accounting for local field effects to model the dielectric function of metal oxides. Furthermore, we reckon the materials energy loss functions $\mbox{Im} (-1/\bar{\epsilon})$, including the accurate evaluation of the momentum transfer dispersion from first-principles. In this respect, by using a Mermin-type parametrization we are able to model the contribution of different electronic excitations to the dielectric loss function. Finally, from the knowledge of the electron inelastic mean free path, together with the elastic mean free path provided by the relativistic Mott theory, we carry out statistical Monte Carlo (MC) charge transport simulations to reproduce the major features of the reported experimental reflection electron energy loss (REEL) spectra of cerium oxides. The good agreement with REEL experimental data strongly supports our approach based on MC modelling informed by ab initio calculated electronic excitation spectra in a broad range of momentum and energy transfers. } \\

\end{tabular}

 \end{@twocolumnfalse} \vspace{0.6cm}

  ]

\renewcommand*\rmdefault{bch}\normalfont\upshape
\rmfamily
\section*{}
\vspace{-1cm}


\footnotetext{\textit{$^{a}$~European Centre for Theoretical Studies in Nuclear Physics and Related Areas (ECT*-Bruno Kessler Foundation) and Trento Institute for Fundamental Physics and Applications (TIFPA-INFN), Trento, Italy }}
\footnotetext{\textit{$^{b}$~Laboratory of Bio-inspired, Bionic, Nano, Meta Materials \& Mechanics, Department of Civil, Environmental and Mechanical Engineering, University of Trento, Italy}}
\footnotetext{\textit{$^{c}$~Bruno Kessler Foundation, Trento, Italy}}
\footnotetext{\textit{$^{d}$~School of Engineering and Materials Science, Queen Mary University of London, UK}}
\footnotetext{\textit{$^{e}$~Departamento de F{\'i}sica, Centro de Investigaci{\'o}n en {\'O}ptica y Nanof{\'i}sica, Universidad de Murcia, Spain}}
\footnotetext{\textit{$^{f}$~Departament de F{\'i}sica Aplicada, Universitat d'Alacant, Spain}}
\footnotetext{\textit{$^{g}$~Peter the Great St. Petersburg Polytechnic University, Russia}}
\footnotetext{*~E-mail: taioli@ectstar.eu; dapor@ectstar.eu}
\footnotetext{\dag~Electronic Supplementary Information (ESI) available: Supplementary Information contains several figures cited in the text, including band structure, ELF, optical properties of cerium oxides, convergence tests, etc...}


\section{Introduction}
Cerium oxides are inner transition metal oxides with promising applications ranging from fuel cells \cite{cite-key} to catalysis for hydrogen and water splitting \citep{Gunasekar2018, Matz2018}, coating technologies \cite{TANG20112806}, glass‐polishing tools \cite{Feng1504} and electrocromic devices \cite{OZER2001391}, or locomotive industries \cite{KASPAR1999285}. Ceria-based materials have also received increasing attention owing to the higher abundance of cerium among the rare‐earth family elements,
their relative narrow band gap, which makes them active in the visible region of the light spectrum (near violet), and their stability under irradiation as well as their large adsorption capacity. \\
\indent Recently, cerium dioxide (CeO$_2$) and dicerium trioxide (Ce$_2$O$_3$) nanoparticles (NPs) have been proposed also as possible enhancers of the relative biological effectiveness (RBE) in hadrotherapy for cancer treatment \citep{McKinnon2016,BRIGGS20131098}.
In this technique, irradiation by protons or heavier ions is aimed to destroy human cells within well localized tumour regions. The use in radiation therapy of high-$Z$ materials (such as gold, platinum, gadolinium, and iron nanoparticles) concentrated near the tumour region, has been shown to enhance the RBE due to an increase in both direct and indirect damage \cite{Jain,Porcel_2010,doi:10.1259/bjr.20140134,doi:10.1063/1.3589914,Kim_2012,Kuncic_2018,Schuemann_2020,https://doi.org/10.1002/wnan.1656}. 
However, while some of these NPs can be toxic for humans, high-$Z$ ceramic oxides, such as CeO$_2$, might offer a viable biocompatible alternative to enhance localized dose and biodamage in radiotherapy \cite{BRIGGS20131098}. Furthermore, ceria nanoparticles are also being investigated as possible radioprotectors in healthy cells \cite{Swartz}, thus making this material more promising during hadrontherapy.
\\
\indent 
Besides these remarkable applications, the study of the cerium-based oxides CeO$_2$ and Ce$_2$O$_3$ is of fundamental interest to understand the electronic and optical properties of materials that exhibit strong electron correlation affected by the presence of highly localized $4f$ states. Indeed, while in experiments both oxides show insulator behaviours in their ground state \cite{Khalifi2016, Goubin2004,Pauly2017}, Density Functional Theory (DFT) predicts correctly a non-magnetic insulator state only for CeO$_2$, while wrongly a metallic one for Ce$_2$O$_3$ \cite{Skorodumova2001,Hay2006}. The latter failure represents a known shortcoming of DFT to describe highly-localized orbitals \cite{Fabris2005} in antiferromagnetic materials that exhibit Mott insulating behaviour.
To correct this deficiency and open the fundamental gap, typically one relies on the time-dependent extension of DFT \cite{Aryasetiawan_1998}. 
A different approach that can be followed in order to better describe the localized $f$ states relies on the use of hybrid functionals \citep{Hay2006, Graciani2011, Brugnoli2018}.
Finally, a more phenomenological way to correct DFT failure in this respect is based on the so-called DFT+U method \cite{Anisimov1997} (or also  on the GW+U approximation \citep{Jiang2009, Jiang2012}, where the GW correction \cite{Aryasetiawan_1998,umari2012communication,taioli2009electronic} is applied on top of LDA+U), in which the addition of an Hubbard correction increases the effective Coulomb repulsion $U_\mathrm{eff}= U-J$ by the sum of the strong on-site Coulomb interaction of localized electrons $U$ and the strength of the exchange interaction $J$ \cite{Dudarev1998}. Within this framework, the localization of the $4f$ states in CeO$_2$ and Ce$_2$O$_3$ can be achieved, resulting in band gap opening. In this regard, several works can be found reporting first-principles simulations of the structural, electronic and optical properties of cerium oxides \citep{Skorodumova2001, Goubin2004, Aguiar2012, Khalifi2016}. However, to the best of our knowledge, significant differences between simulations and experimental results persist, particularly in the assessment of the optical properties.
\\
\indent 
Here we focus our study on the ab initio calculation of the electronic band structure and of the dielectric response of bulk Ce({\scriptsize +IV})O$_2$ and Ce({\scriptsize +III})$_2$O$_3$, which are characterised by the two known oxidation states of cerium in solids, using the time-dependent Density Functional Theory method (TDDFT) \citep{Runge1984} in linear response (LR) approximation \citep{Botti2007}. We adopt an interacting all-electron approach not plagued by the presence of pseudopotentials (that deal implicitly with core and  explicitly with chemically active valence states), the only approximation being the interaction potential among electrons.
The macroscopic dielectric function $\bar{\epsilon}({\bf q},W)$, or the closely related energy loss function ELF$({\bf q},W)$, where ${\bf q}$ and $W$ are the momentum and energy transfers, respectively, are used to describe the materials excitation spectrum and, thus, the inelastic interactions between charged particles and the medium constituents \cite{Lindhard1954,Ritchie1959}. 
In particular, we show that the inclusion of local field effects (LFE) overcomes the discrepancy between simulations and experimental measurements.
\\
\indent The ELF, conveniently weighted and integrated, provides the electron inelastic scattering cross section \cite{Ritchie1957,Ritchie1959}, which is used, along with the elastic scattering cross section derived from the relativistic Mott theory \cite{Mott1929}, as input to a Monte Carlo (MC) routine to model the transport of charged particles within these solids and predict the reflection electron energy loss (REEL) spectra in particular~\cite{Dapor2020book,azzolini2020comparison,azzolini2018secondary,azzolini2018anisotropic,azzolini2017monte,taioli2015computational,taioli2009mixed,taioli2010electron}.
REEL spectroscopy is an analysis technique that uses an electron beam impinging with kinetic energy lower than 2~keV into thin films. Primary electrons penetrate a few nanometers into the material surface, lose their energy via inelastic collisions and are eventually backscattered to the spectrometer \cite{egerton}, resulting in spectra that are typically characterised by a number of structures attributed to both collective (plasmons) and single-electron excitations and can be directly benchmarked against available experimental data.
Here, the agreement of the simulated and the measured REELS is used to validate the TDDFT-calculated electronic excitation spectra of cerium oxides in a broad energy and momentum transfers region.\\
\indent 
This paper is organized as follows. In section 2 we summarize the theoretical methods used to perform the ab initio simulations of the dielectric function and the MC strategy. In section 3 we discuss first the electronic and optical properties of cerium oxides, such as the ELF in momentum-energy space. Furthermore, a detailed analysis of the excitation spectra is carried out using the Mermin Energy Loss Function--Generalized Oscillator Strength (MELF--GOS) approach \cite{PhysRevA.58.357,PhysRevA.72.052902,Garcia-Molina2012,https://doi.org/10.1002/sia.5947}. Finally, the REEL spectra are obtained by applying our MC method and compared with experimental measurements. Atomic units are used throughout this work except otherwise stated.

\section{Theoretical and computational methods}
\label{sec:ComputationalDetails}
Here we provide the fundamentals of the methodology used to obtain the atomic structures and the excitation spectra of bulk CeO$_2$ and Ce$_2$O$_3$. These properties are required to obtain reliable cross sections needed to simulate the main interactions that affect the electron transport through these materials.  

\subsection{Materials structure}
\label{subsec:MaterialsStructure}
The DFT+U method \cite{Anisimov1997} has been employed to optimize both the cell volume and atomic positions of the cerium oxides. 
On the one hand, CeO$_2$ is characterised by a face centered cubic fluorite-type structure, belonging to the $Fm\overline{3}m$ point group. The primitive cell of CeO$_2$ (inset in Fig. \ref{fig:PDOS}) has a crystal basis defined by two oxygen atoms, which in lattice coordinates are positioned at (0.25, 0.25, 0.25) and (0.75, 0.75, 0.75), respectively, and one cerium atom, centered in (0, 0, 0). 

On the other hand, Ce$_2$O$_3$ has a trigonal geometry, belonging to the $P3\overline{m}1$ point group. The primitive cell (inset in Fig. \ref{fig:DOSCe2}) includes three oxygen atoms at the crystallographic positions (0, 0, 0), (1/3, 2/3, 0.642), and (2/3, 1/3, 0.3574) in lattice coordinates, respectively, and two cerium atoms centered at (1/3, 2/3, 0.2459) and (1/3, 2/3, 0.754).

We report in Table \ref{tbl:Cells} the lattice constants of the optimized cells. The motivations to fix the value of $U_\mathrm{eff}=5$~eV, which delivers a good agreement with experimental data, will be discussed further below.

\begin{table}[htbp]
\centering
\small
\caption{Calculated CeO$_2$ and Ce$_2$O$_3$ cell parameters in comparison with experimental results.}
\label{tbl:Cells}

\begin{tabular}{lcccccc} \hline
Method  & $U_\mathrm{eff}$ &  CeO$_2$ & \multicolumn{2}{c}{Ce$_2$O$_3$}   &  Ref.       \\
       &  & $a$ (\AA)    & $a$ (\AA) &  $c$ (\AA)  &        \\  \hline
LSDA+U         &   5    &    5.36        &  3.83  &   5.95 &  This work                \\
Experim.          &        &     5.411      &          &      & \citep{GERWARD2005}   \\
Experim.          &        &  &   3.891  &   6.059  & \citep{BARNIGHAUSEN1985} \\   \hline
\end{tabular}
\end{table}

\subsection{Dielectric response }
\label{subsec:DielectricResponse}
The materials microscopic dielectric function depends on the bare Coulomb potential $v_{\rm C}$  and the polarization function $\chi({\bf q},W)$ through \cite{RevModPhys.74.601}:
\begin{equation}
\epsilon({\bf q},W)=1-v_{\rm C}({\bf q})\chi({\bf q},W) \, \mbox{.}
\end{equation}
The polarization function can be obtained by solving the following Dyson-like equation:
\begin{equation}\label{Dyson}
    \chi^{-1}({\bf q},W)=\chi^{-1}_{0}({\bf q},W)-v_{\rm C}({\bf q})- f_{\rm xc}({\bf q},W) \, \mbox{,}
\end{equation}
where $\chi^{-1}_{0}({\bf q},W)$ is the non-interacting (or independent-particle) polarization calculated from the Kohn-Sham wavefunctions and $f_{\rm xc}({\bf q},W)$ is the TDDFT kernel.
In this regard, we have used the adiabatic local density approximation (ALDA) kernel, which is related to the LDA exchange-correlation functional $v_{\rm  xc}[\rho]$ by:
\begin{equation}
    f_{\rm xc}(\textbf{r},t)= \left\{ \frac{d}{d\rho}v_{\rm  xc}[\rho]\right\}_{\rho=\rho(\textbf{r},t)},
\end{equation}
where $\rho$ is the DFT ground state density.
This approach is supported by previous works on inelastic X-ray scattering (IXS) for finite momentum transfer calculations where the ALDA kernel showed a good agreement with the experimental findings due to correct inclusion of short range terms \cite{PhysRevB.93.035128,PhysRevB.81.085104,PhysRevB.97.125144,ParedesMellone2020}.\\ 
\indent For periodic crystals one can exploit the translational symmetry, and the microscopic dielectric function can be conveniently written in reciprocal space, i.e. $\epsilon_{\mathbf{G},\mathbf{G'}}(\mathbf{q}, W) = \epsilon(\mathbf{q}+\mathbf{G}, \mathbf{q}+\mathbf{G'}, W)$, where $\mathbf{G}$ and $\mathbf{G'}$ are reciprocal lattice vectors, and $\mathbf{q}$ is the transferred momentum vector in the first Brillouin zone (1BZ). Using this notation, $\epsilon_{\mathbf{G},\mathbf{G'}}(\mathbf{q}, W)$ is also often called the dielectric matrix. It can be shown \cite{Adler,Wiser} that the experimentally measurable macroscopic dielectric function $\bar{\epsilon}$ and the microscopic one $\epsilon$ are related by the following expression:
\begin{equation}\label{eq:eM}
\bar{\epsilon}( \mathbf{q}, W) =  \left [ \epsilon^{-1}_{\mathbf{G}=0,\mathbf{G'}=0}(\mathbf{q}, W)  \right ]^{-1}.
\end{equation}
Local field effects (LFE), which are related to local density inhomogeneities, are included by  inverting the full dielectric matrix and taking subsequently the head of the inverse matrix. The effects of local density inhomogeneities in the dielectric response of a material are indeed embedded in the wings of the microscopic dielectric matrix. \\
\indent The dielectric response function has been obtained using the LR-TDDFT implementation of the ELK code suite \cite{ELK} on top of the ground state obtained by the DFT+U method. ELK uses an all-electron Full-Potential Linearized Augmented-Plane-Wave (FP-LAPW) approach. The Hubbard correction has been treated in the fully localised limit (FLL). To deal with the antiferromagneticity of Ce$_2$O$_3$, we included both spin polarization and spin-orbit coupling. The local spin density approximation (LSDA) exchange correlation functional \cite{Perdew1992} has been used for the ground state calculations alongside the ALDA approximation for the time dependent exchange correlation functional. We also checked the convergence with respect to the number of $k$-points in the 1BZ, finding well converged results for a 10 $\times$ 10 $\times$ 10 $k$-point grid for CeO$_2$ and for a 8 $\times$ 8 $\times$ 8 $k$-point grid for Ce$_2$O$_3$. A number of empty bands equal to $50$ for each atom have been used to obtain converged results up to $120$~eV. \\
\indent To perform the extension of the  dielectric function in the optical limit (i.e. ${\bf q}=0)$ to finite momenta the transferred momentum has been selected along the [111], [110], [211] directions in reciprocal space for CeO$_2$ and along the [001], [110], [100] directions for Ce$_2$O$_3$, respectively, with modulus set to $6$ \AA$^{-1}$ in the reciprocal space.

\subsection{Energy loss function}
\label{subsec:ELF}
To describe the propagation of charged particles through matter using the dielectric formalism  \cite{Lindhard1954,Ritchie1957,Ritchie1959}, one relies on the material energy loss function  ELF, which is related to the macroscopic dielectric function as follows:
\begin{equation}\label{ELF}
\text{ELF} = \text{Im} \left [\frac{-1}{\bar{\epsilon}(\mathbf{q}, W)} \right ].
\end{equation} 

In principle, to determine the inelastic scattering cross section one needs to know the dependence of the ELF over the entire spectrum of meaningful excitation energies $W$ and momentum transfer ${\mathbf{q}}$ \cite{azzolini2017monte,taioli2020relative,azzolini2020comparison,Garcia-Molina2012}. However, typically one has access only to a limited range of energies, corresponding to those of the valence electrons ($\lesssim 100$~eV), owing to the prohibitive computational effort of including high-energy excitations as well as their momentum dispersion. \\
\indent Thus, to extend the excitation energy range beyond the valence regime, we propose to use the MELF--GOS model \cite{PhysRevA.58.357,PhysRevA.72.052902,Garcia-Molina2012,https://doi.org/10.1002/sia.5947}, which implements a numerically effective and accurate method to compute the ELF over the entire Bethe surface (i.e., the momentum and energy transfers plane) by including both valence and inner shell electronic excitations. 
In short, the derivation of the ELF within the MELF--GOS framework is based on a fitting procedure, whereby the optical-limit ab initio (or measured) energy-loss spectrum of the target material is equated to a zero-momentum-transfer functional form that for the valence (outer) contribution can be written as a sum of Mermin-type ELFs \cite{PhysRevB.1.2362,doi:10.1021/acs.jpcc.8b10832}:
\begin{eqnarray}\label{mermin} 
&&\text{Im}\Big[ \frac{-1}{\bar{\epsilon}({\bf{q}=0},W)}\Big]_{\rm outer} = \nonumber \\ && \sum_i F(W-W_{\mathrm{th},i}) 
\text{Im}\Big[ \frac{-1}{\epsilon_{\rm M}(A_i,W_i,\gamma_i;{\bf{q}=0},W)} \Big],
\end{eqnarray} 
where:
\begin{equation}\label{Drude}
\text{Im}\Big[ \frac{-1}{\epsilon_{\rm M}(A_i,W_i,\gamma_i;{\bf{q}=0},W)} \Big] = \frac{A_i \gamma_i W}{(W_i^2-W^2)^2+(\gamma_i  W)^2},   
\end{equation}
and
\begin{equation}\label{smooth}
F(W-W_{\mathrm{th},i})=\frac{1}{1+e^{-\Delta_i
(W-W_{\mathrm{th},i})}}
\end{equation}
is a smooth switching function characterised by steepness $\Delta_i$, which is used as an additional fitting parameter to soften the onset of the outer-shell electronic excitations at the threshold energies $W_{{\rm th},i}$ ($F$ can also be represented by a Heaviside step function). 
The other fitting parameters in eqn (\eqref{mermin}), $A_i$, $W_i$, and $\gamma_i$, are related, respectively, to the relative weight, position, and width of the peaks observed in the optical ELF spectrum; they are determined so as to reproduce the main features of the ELF obtained from ab initio calculations.
Within the MELF--GOS framework one then makes an interpolation of the TDDFT ab initio data to higher excitation energies, using an $a W^{-b}$ functional form, where $a$ is a constant and $b$ is a fit parameter obtained from the ab initio data. 
The contribution of inner-shell electrons is included by means of atomic generalized oscillator strengths (GOS) as follows \cite{https://doi.org/10.1002/sia.5947}:
\begin{equation}\label{mermin2}
\text{Im}\Big[ \frac{-1}{\bar{\epsilon}({\bf q}=0,W)}\Big]_{\rm inner} = \frac{2\pi^2 {\cal N}}{W}\sum_j \alpha_j \sum_{nl} \frac{df_{nl}^j({\bf q}, W)}{d W} \Theta(W - W_{{\rm th},nl}^{j}),
\end{equation}
where $\frac{df_{nl}^j({\bf q}, W)}{d W}$ are the GOS of the hydrogenic wavefunctions that are obtained using an effective nuclear charge for each inner shell identified by the atomic quantum numbers $(n,l)$ of the target $j$-th constituent with stoichiometric weight $\alpha_j$. In eqn (\ref{mermin2}), $W_{{\rm th},nl}^{j}$ is the ionization energy of the orbital, and ${\cal N}$ is the target atomic or molecular number density.

\subsection{Monte Carlo simulations of electron transport}
\label{subsec:MCsimulationREELS}
In the MC approach the interaction between the incident electron beam and a cerium oxide target is simulated by treating the electrons as point particles that follow classical trajectories within the material. This is a viable representation at high enough energy \cite{LILJEQUIST201445}. Electron trajectories are determined by the elastic and inelastic interactions, which are assessed using quantum mechanical methods. The occurrence
of either an elastic or inelastic scattering event at kinetic energy $T$ is assessed by comparing a random number uniformly distributed in the range $[0,1]$ with the relevant probabilities $p_\mathrm{el} (T)={\Lambda_\mathrm{el}} (T)/{\Lambda_{\mathrm{tot}}} (T)$ and $p_\mathrm{inel} (T)= {\Lambda_\mathrm{inel}} (T)/{\Lambda_{\mathrm{tot}}}(T)$, where $\Lambda_\mathrm{tot}(T) = \Lambda_\mathrm{inel}(T) + \Lambda_\mathrm{el}(T)$.\cite{Dapor2020book} $\Lambda_{\rm el/inel}=\lambda_{\rm el/inel}^{-1}(T)$ 
is the inverse mean free path for the elastic or inelastic scattering. 
\\
\indent Once the type of collision has been chosen, the angular deviation of the electron trajectory after an elastic collision, given by the scattering angle $\theta$, can be evaluated by equating the elastic scattering cumulative probability
\begin{equation}
    \label{elcum}
    P_\mathrm{el}(T, {{\theta}}) = \frac{2\pi}{\Lambda_\mathrm{el}(T)} \int_0^{{\theta}}  \frac{d\Lambda_\mathrm{el}(T,\theta')}{d\Omega} \sin\theta' d\theta',
\end{equation}
to a uniformly-distributed random number generated in the range $[0,1]$. No energy loss is assumed in an elastic collision.
\\
\indent In a similar way, the energy loss $W$ due to an inelastic scattering event of an electron with kinetic energy $T$ can be reckoned by equating the inelastic scattering cumulative probability distribution:
\begin{equation}
\label{inelcum}
P_\mathrm{inel}(T,W) = \frac{1}{\Lambda_\mathrm{inel}(T)} \int_{0}^{W} \frac{d \Lambda_\mathrm{inel}(T,W')}{dW'}dW' 
\end{equation}
to another uniformly-distributed random number in the range $[0, 1]$. The angular deviation due to inelastic scattering is evaluated according to the classical binary collision theory\cite{Dapor2020book}.\\
\indent The main quantity to account for inelastic events that electrons with kinetic energy $T$ undergo when moving through a medium characterised by an ELF is the differential inverse inelastic mean free path (DIIMFP):
\begin{equation}
\frac{d\Lambda_\mathrm{inel}(T,W)}{dW} = \frac{1}{\pi T} \int_{q_-}^{q_+} \frac{1}{q} \text{Im} \left [ \frac{-1}{\bar{\epsilon}(\mathbf{q}, W)} \right ]dq, 
\label{EqDIIMFP}
\end{equation}
where the integration limits
$q_{\pm} = \sqrt{2T} \pm  \sqrt{2(T-W)}$,
result from energy and
momentum conservation in the interaction process.\\
\indent From the knowledge of the DIIMFP one can reckon 
the IMFP to be used in the MC simulation of electron transport through the cerium oxides by calculating the following inelastic scattering cross section:
\begin{equation}
\Lambda_\mathrm{inel}(T) = \int_{W_\mathrm{min}}^{W_\mathrm{max}} \frac{d\Lambda_\mathrm{inel}(T,W)}{dW} dW,
\label{EqIMFP}
\end{equation}
where the integration limit $W_\mathrm{min}$ is set to $E_\mathrm{gap}$ for semiconductors and insulating materials, while $W_\mathrm{max}$ represents the minimum between $T$ and $(T+W_\mathrm{min})/2$. 
By inverting eqn (\ref{EqIMFP}) one can obtain the inelastic mean free path (IMFP), which represents the average distance between inelastic collisions.
The effect on the IMFP of electron-polaron and electron-phonon scattering events will be neglected due to the high kinetic energy ($T \simeq 1.6$~keV) considered in this work. \\
\indent The elastic scattering is accounted for by means of the Mott theory \cite{Mott1929}, which generalizes the screened Rutherford scattering model also at low kinetic energy of the colliders and for high-$Z$ materials. This approach provides the differential elastic scattering cross-section (DESCS) for an electron being scattered at an angle $\theta$ when impinging on a central potential as follows \cite{Dapor2020book}:
\begin{equation}\label{scatel}
\frac{d\Lambda_{\mathrm{el}}(T,\theta)}{d\Omega}\,={\cal N}\,[|f(\theta)|^2+|g(\theta)|^2],
\end{equation}
where $f(\theta)$ and $g(\theta)$ are the direct and spin-flip scattering amplitudes, respectively, which can be obtained by solving the Dirac equation in a central field. This observable represents the probability per unit solid angle that an electron is scattered by the target centers. \\
\indent However, in condensed matter the central symmetry of atomic systems is lost, and thus we must generalize to some extent the Mott theory to take into account the presence of bonded interactions among neighbours in the periodic unit cell.
Condensed phase effects can be achieved by allowing interference between the direct  $f_m(\theta)$ and spin-flip $g_m(\theta)$ scattering amplitudes of the $m^{\mathrm{th}}$--atom and those of the neighbouring $n^{\mathrm{th}}$--atom. These interference terms account for multiple scattering events among the scattered waves modifying eqn (\ref{scatel}) as follows: 
\begin{equation}\label{molsca}
\frac{d\Lambda_{\mathrm{el}}(T,\theta)}{d\Omega}\,={\cal N}\,\sum_{m,n}\,\exp(i {\bf q} \cdot {\bf r}_{mn})\,[f_m(\theta) f^*_n(\theta)\,+\,g_m(\theta) g^*_n(\theta)]\,,
\end{equation}
where ${\bf r}_{mn}\,=\,{\bf r}_m\,-\,{\bf r}_n$, and ${\bf r}_m~({\bf r}_n)$ identifies the position of the $m^{\mathrm{th}}~(n^{\mathrm{th}})$ atom in the periodic unit cell. \\
\indent 
Finally, by integrating over the solid angle one obtains the total elastic scattering cross section:
\begin{equation}
 \Lambda_{\mathrm{el}}(T) = \int_{\Omega} \frac{d \Lambda_{\mathrm{el}}(T,\theta)}{d \Omega} {d \Omega}.
\end{equation}


\section{Results and discussion}

In the following, the main features of the electronic excitation spectra of CeO$_2$ and Ce$_2$O$_3$ will be discussed. First, the band structure, the total and partial density of states (DOS) of electrons will be analyzed, 
which serves to set up the Hubbard parameter $U_{\rm eff}$ for the TDDFT calculations. These DOS will be then useful to interpret the dielectric properties of cerium oxides. 
Finally, the electron inelastic scattering cross sections will be obtained 
from the complex dielectric functions, and will be used to perform the MC simulations of REEL spectra.

\begin{figure}[t]
\centering
\includegraphics[width=0.9\columnwidth]{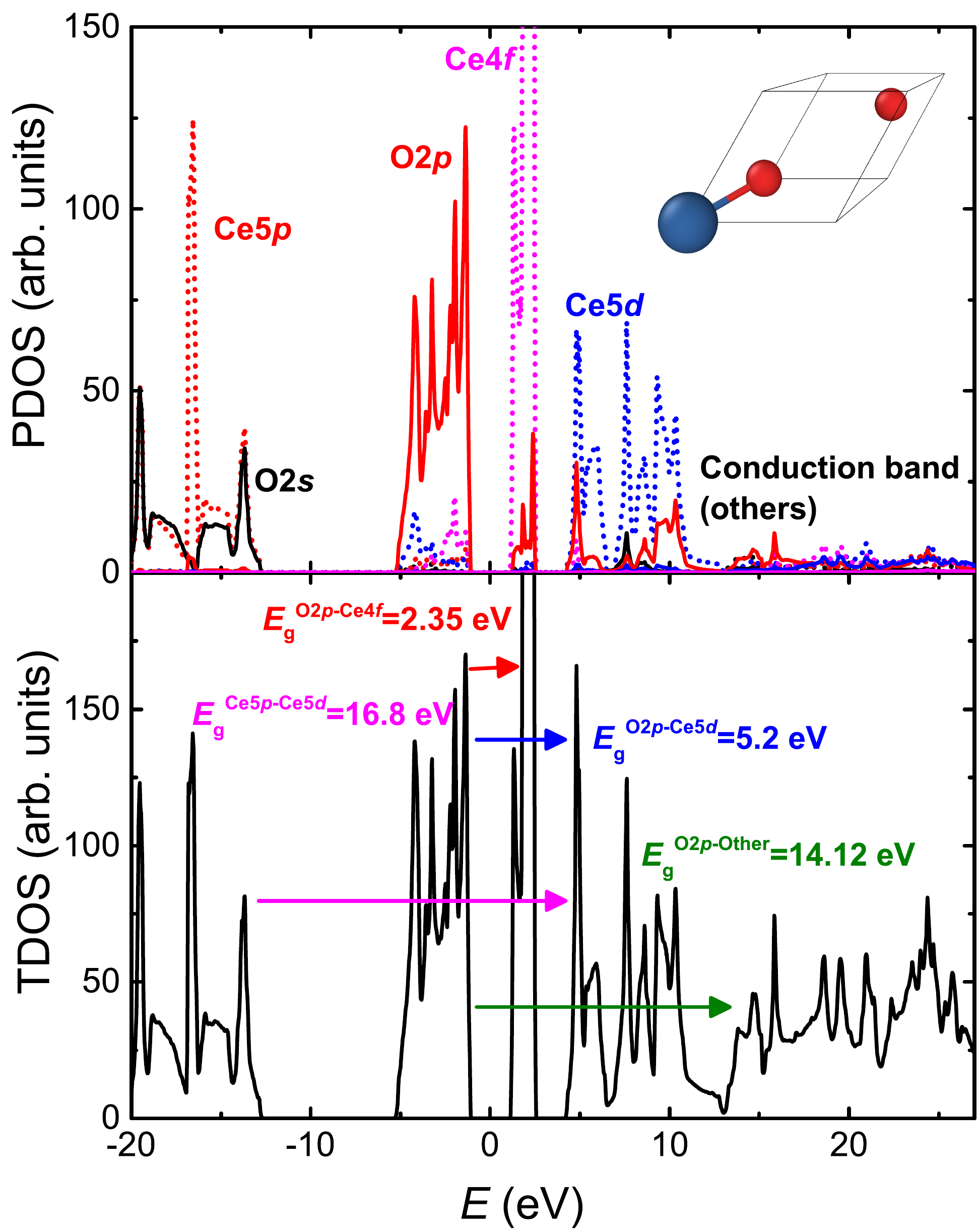}
\caption{Top panel: partial DOS of bulk CeO$_2$ ($U_\mathrm{eff}=5$~eV) showing orbitals contributions for both cerium (dashed lines) and oxygen (continuous lines). Bottom panel: Total DOS. The Fermi level is set to the origin of the energy axis. Inset: CeO$_2$ primitive cell, cerium and oxygen in blue and red color, respectively.}
\label{fig:PDOS}
\end{figure}

\subsection{Band structures and DOS of bulk CeO$_2$ and Ce$_2$O$_3$}
\label{sec:DOS}

\begin{figure}[t]
\includegraphics[width=0.9\columnwidth]{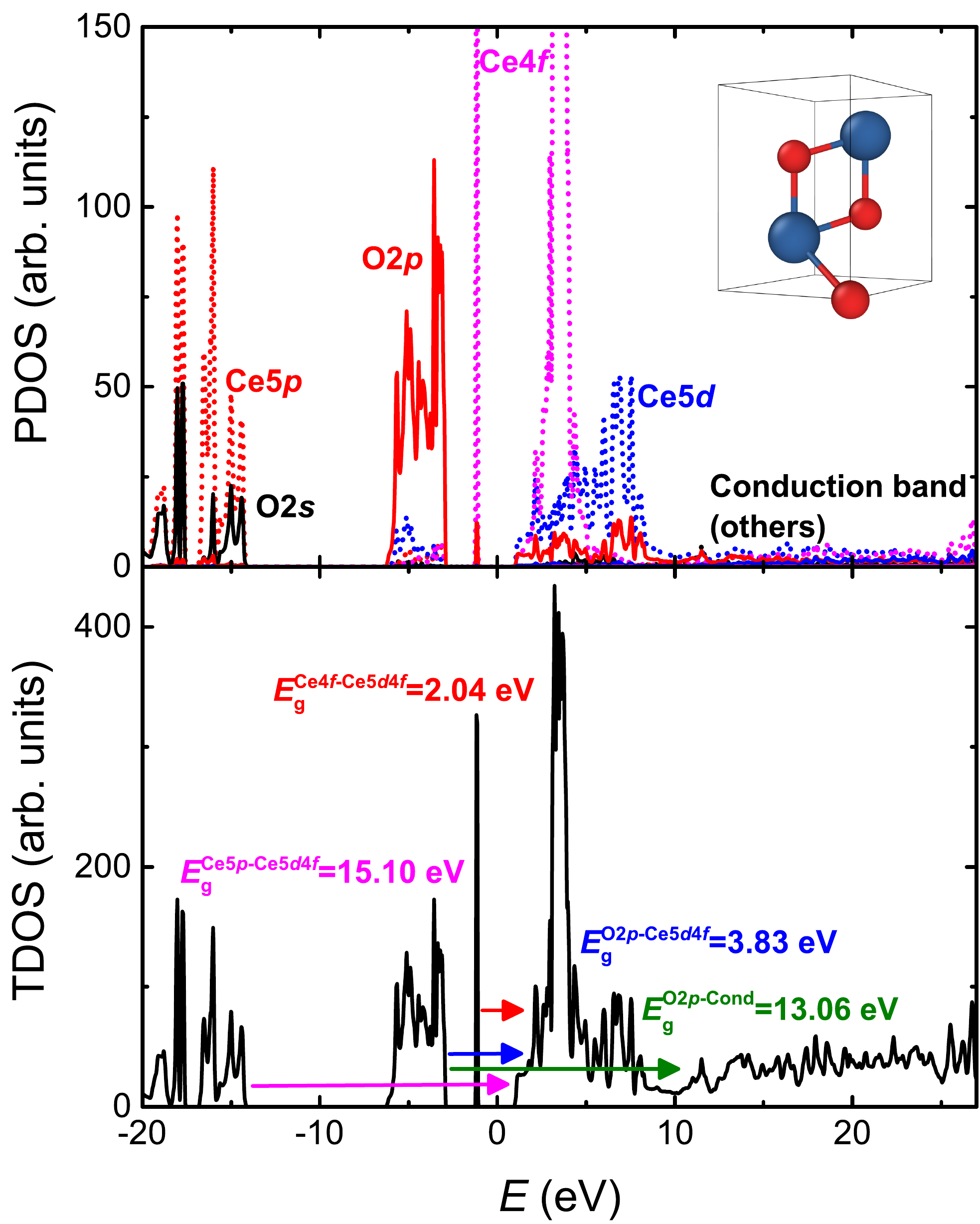}
\caption{Top panel: partial DOS of bulk Ce$_2$O$_3$ ($U_\mathrm{eff}=5$~eV) showing orbitals contributions for both cerium (dashed lines) and oxygen (continuous lines). Bottom panel: Total DOS. The Fermi level is set to the origin of the energy axis.  Inset: Ce$_2$O$_3$ primitive cell, cerium and oxygen in blue and red color, respectively.}
\label{fig:DOSCe2}
\end{figure}

Ground state DFT calculations have been carried out, aiming particularly to check the dependence of the electronic band structure on different values of the Hubbard correction $U_\mathrm{eff}$. 
After a number of tests carried out on the total density of states (DOS) (see Figs. S1 and S2 of the Supplementary Information) and on the ELF in the optical limit (see Fig. S5 of the Supplementary Information), $U_\mathrm{eff}$ was set to 5~eV. This value has been fixed so that i) the localized $4f$ band blueshifts with respect to plane LDA and opens the fundamental gap, particularly in the case of Ce$_2$O$_3$ for which LDA wrongly predicts a metallic character; ii) the optical ELF calculated with this value of $U_\mathrm{eff}$ is in fair agreement with the experimental data available for CeO$_2$.\\
\indent In CeO$_2$ the fundamental gap occurs between the top of the valence band with O$2p$ character and the bottom of the conduction band formed by the localized Ce$4f$ orbitals. This can be observed in the bottom panel of Fig. \ref{fig:PDOS}, where we plot the total DOS of bulk CeO$_2$ along with the relevant band gaps  (the Fermi level is set to the origin of the energy axis). In particular, in the top panel of Fig. \ref{fig:PDOS} we show the contribution to the total DOS of cerium (dashed lines in the top panel) and oxygen (continuous lines in the top panel), respectively, along with the projection on the relevant angular momenta (PDOS).
This analysis shows clearly the hybridization between different symmetry orbitals of O and Ce.
In particular, a deeper occupied band (below $-10$~eV in Fig.  \ref{fig:PDOS}) is mainly formed by Ce$5p$ mixed with O$2s$, while O$2p$ and Ce$4f$ hybridize between $-5$~eV and 0~eV. Above the valence band, between 0~eV and 5~eV, there is a localized Ce$4f$-O$2p$ narrow band and beyond that we find a conduction band mainly of Ce$5d$ nature with a mixture of several O and Ce levels at larger energies between 5~eV and 15~eV (see also Fig. S3 of the Supplementary Information). 

\begin{table}[b]
\small
\caption{\ Band gap opening (eV) in bulk CeO$_2$ using different exchange-correlation functionals in DFT simulations.}
 \label{tbl:BANDGAP1}
 \begin{tabular*}{0.48\textwidth}{@{\extracolsep{\fill}}lllll}
    \hline
Method  & $U_\mathrm{eff}$ & Band gap & Band gap &  Ref.  \\
 & &   O$2p\rightarrow$Ce$4f$    &  O$2p \rightarrow$Ce$5d$ &               \\ \hline
LSDA          &  &    2.05      &  5.8    &   This work                \\
  
LSDA+U      &  3 &  2.35 &      5.5             &   This work       \\

LSDA+U   &  5  &   2.35   &      5.3   &   This work                \\

GGA           &  &    2.3      &          &   \citep{Skorodumova2001}  \\

PBE           &    & 1.7       &          &    \citep{Jiang2005}       \\

PBE           &  &   1.7, 1.9  &          &   \citep{Hay2006}          \\

PBE    &   &  1.8       &   5.7    &    \citep{Yang2004}        \\

HSE       &  &  3.5, 3.3   &   7.0    &  \citep{Hay2006}            \\

Experim.   &  &  3           &   6–8   &   \citep{Wuilloud1984}      \\
    \hline
  \end{tabular*} 
\end{table}

\begin{table}[b]
\small
  \caption{\ Gaps between bands (eV) in bulk Ce$_2$O$_3$ using different $U_{\rm{eff}}$ in DFT simulations.}
\label{tbl:BANDGAP2}
  \begin{tabular*}{0.48\textwidth}{@{\extracolsep{\fill}}lllll}
    \hline
    Method  & $U_\mathrm{eff}$ & Band gap   & Band gap & Band gap \\
 & &   O$2p\rightarrow$Ce$4f$    &  O$2p\rightarrow$Ce$5d$ &  Ce4$f\rightarrow$Ce$4f,5d$                                        \\
    \hline
 LSDA+U         &   5    &    1.1        &     3.95   &   2.04               \\

LSDA+U        &    6   &     1.7       &    3.97     &  2.59              \\

    \hline
  \end{tabular*}
\end{table}

In Table \ref{tbl:BANDGAP1} we also report, in comparison with other theoretical and experimental data, the gaps between the O$2p$ valence band and the $4f$ states and between the O$2p$ valence band and the Ce$5d$ conduction band with the exclusion of the $4f$ states. The latter is more affected by the fine tuning of $U_\mathrm{eff}$. 
Plain DFT values show the typical underestimation of the fundamental gap (2.05~eV) with respect to experimental measurements (ranging from 3.3 to 3.6~eV \cite{doi:10.1063/1.359225,PFAU199471,PhysRevLett.106.246801}), slightly improved to 2.35~eV by adding the Hubbard correction $U_\mathrm{eff}$.\\ 
\indent 
The full band structure of CeO$_2$ calculated using DFT+U is given in Fig. S3 of the Supplementary Information for reference. Both conduction (above 4~eV) and valence (below $-1$ eV) bands show some significant dispersion, while the $4f$ states between 0.8 and 2~eV have smaller bandwidth. From the band plot analysis it turns out that CeO$_2$ is an indirect gap insulator.
Our results are in very good agreement with previous studies using hybrid functionals \cite{Hay2006,Khalifi2016}.\\
\indent In the case of Ce$_2$O$_3$, the addition of the Hubbard correction is critical in the simulation of the electronic band structure and DOS 
as this material is wrongly predicted to be a metal by DFT-LDA. In fact, the Ce atom in Ce$_2$O$_3$ displays a $+3$ oxidation state, where one electron is expected to populate the $4f$ band leading to possible ferromagnetic or antiferromagnetic ground state. By switching on the Hubbard correction we obtain the localization of the $4f$ states, with redshift of the latter bands by increasing $U_\mathrm{eff}$ from $5$ to $6$~eV (see a plot of the DOS of bulk Ce$_2$O$_3$ for $U_\mathrm{eff}=5, 6$~eV in Fig. S2 of the Supplementary Information) and with blueshift of the conduction bands at higher energies.
The gaps between different bands are reported in Table \ref{tbl:BANDGAP2}.\\
\indent In Fig. \ref{fig:DOSCe2} we show the total (TDOS, bottom panel) and projected (PDOS, top panel) DOS of Ce$_2$O$_3$. The full band structure  calculated using DFT+U is given in Fig. S4 of the Supplementary Information for reference. From the analysis of the PDOS we notice that the Ce$4f$ band splits in two subbands, a very narrow occupied band, and a narrow and intense empty band that overlaps with the empty Ce$5d$ band above the Fermi level (the origin of the energy axis), respectively. The DOS is also characterised by a valence band mainly of O$2p$ nature, a deeper occupied band mainly of Ce$5p$ origin, and an outer conduction band
above 10~eV which is a mixture of high energy states from all constituents. The almost localized nature of the $4f$ band can be seen in 
Fig. S4 of the Supplementary Information, where a flat band around $-1$~eV from the Fermi level appears. At variance, the other conduction bands show a large dispersion owing to the hybridization of the Ce$4f,5d$ states with O$2p$ orbitals. Finally, we notice that also Ce$_2$O$_3$ is an indirect gap insulator, being found the top of the valence band at the $\Gamma$-point and the bottom of the conduction band at the $K$-point.


\subsection{The excitation spectrum of bulk CeO$_2$}
\label{sec:CeO2}
The excitation spectrum of bulk materials is determined by the real ($\bar{\epsilon}_1$) and imaginary ($\bar{\epsilon}_2$) parts of the complex macroscopic dielectric function (see eqn (\ref{eq:eM})). The former informs about the screening capacity of the material to electric fields, while the latter provides the spectrum of possible electronic transitions between occupied and empty states. They are directly connected to the optical refractive index and extinction coefficient (see Fig. S10 and eqn (S1) of the Supplementary Information), so they can be straightforwardly checked against experimental optical determinations for ${\bf{q}} \rightarrow 0$. Particularly, $\bar{\epsilon}_2$ can be correlated with the PDOS obtained in subsection \ref{sec:DOS}, which allows to identify the origin of the different excitations. The latter are in turn connected to the structure of the ELF, and this identification may be relevant for the assessment of the quality of the obtained spectra (see sections 3.2 and 3.3 below) and the calculation of the inelastic mean free path \cite{doi:10.1021/acs.jpcc.8b10832}.

\begin{figure}[t]
\centering
	\includegraphics[width=0.9\columnwidth]{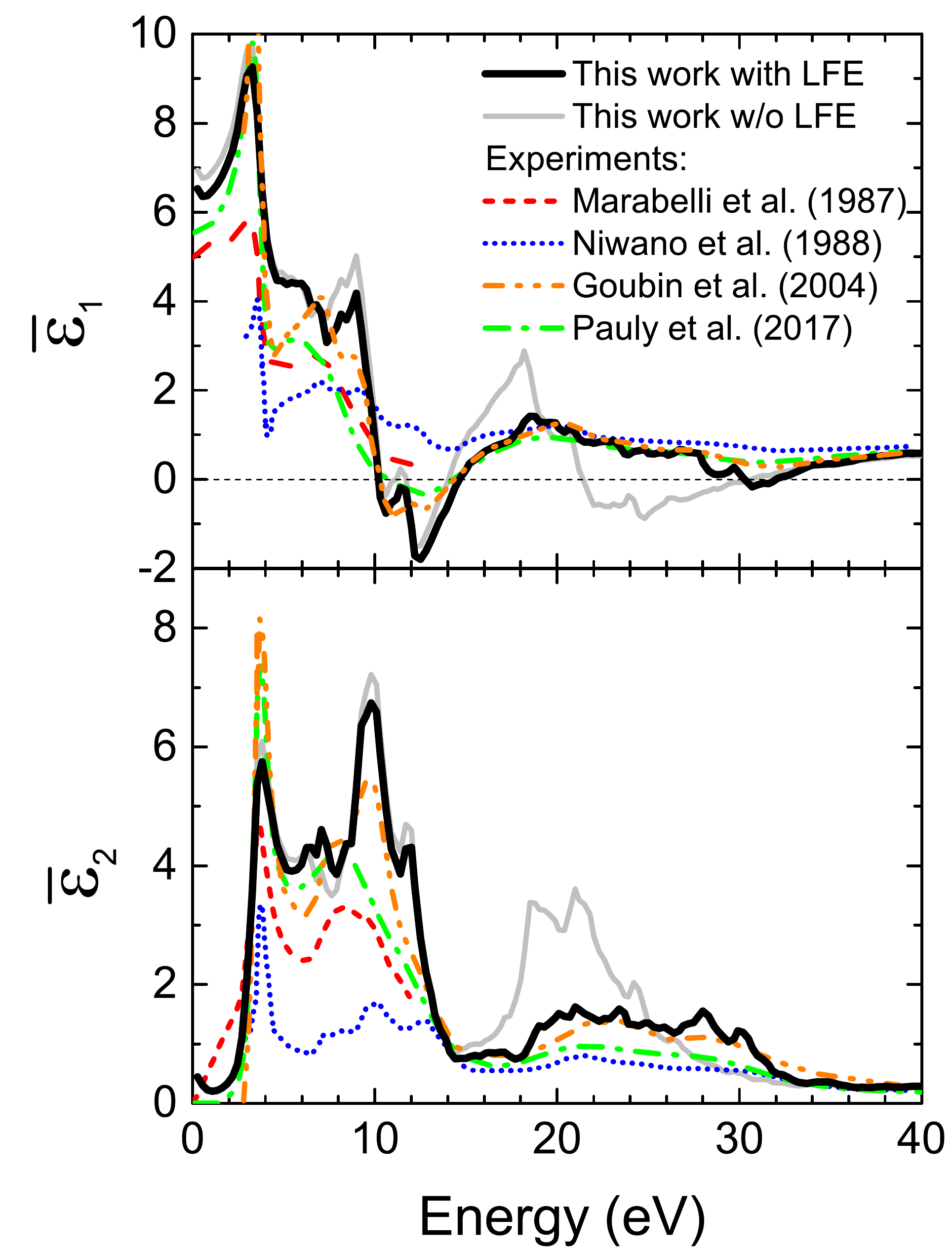}
\caption{Top panel: comparison between the real part ($\bar{\epsilon}_1$) of the experimental dielectric function of bulk CeO$_2$ \citep{Marabelli1987,Niwano1988, Goubin2004,Pauly2017} and our calculation. Bottom panel: comparison between the imaginary part ($\bar{\epsilon}_2$) of the experimental dielectric function  of bulk CeO$_2$ \citep{Marabelli1987,Niwano1988, Goubin2004,Pauly2017} and our calculation. Both $\bar{\epsilon}_1$ and $\bar{\epsilon}_2$ are calculated along the [111] direction of the momentum transfer, taking afterwards the ${\bf{q}} \rightarrow 0$ limit. The inclusion of LFE improves the agreement with experimental data, flattening in particular the features in the range 15--30~eV.}
\label{fig:ComparisonEPS1}
\end{figure}

\begin{figure}[t]
\centering
\includegraphics[width=0.9\columnwidth]{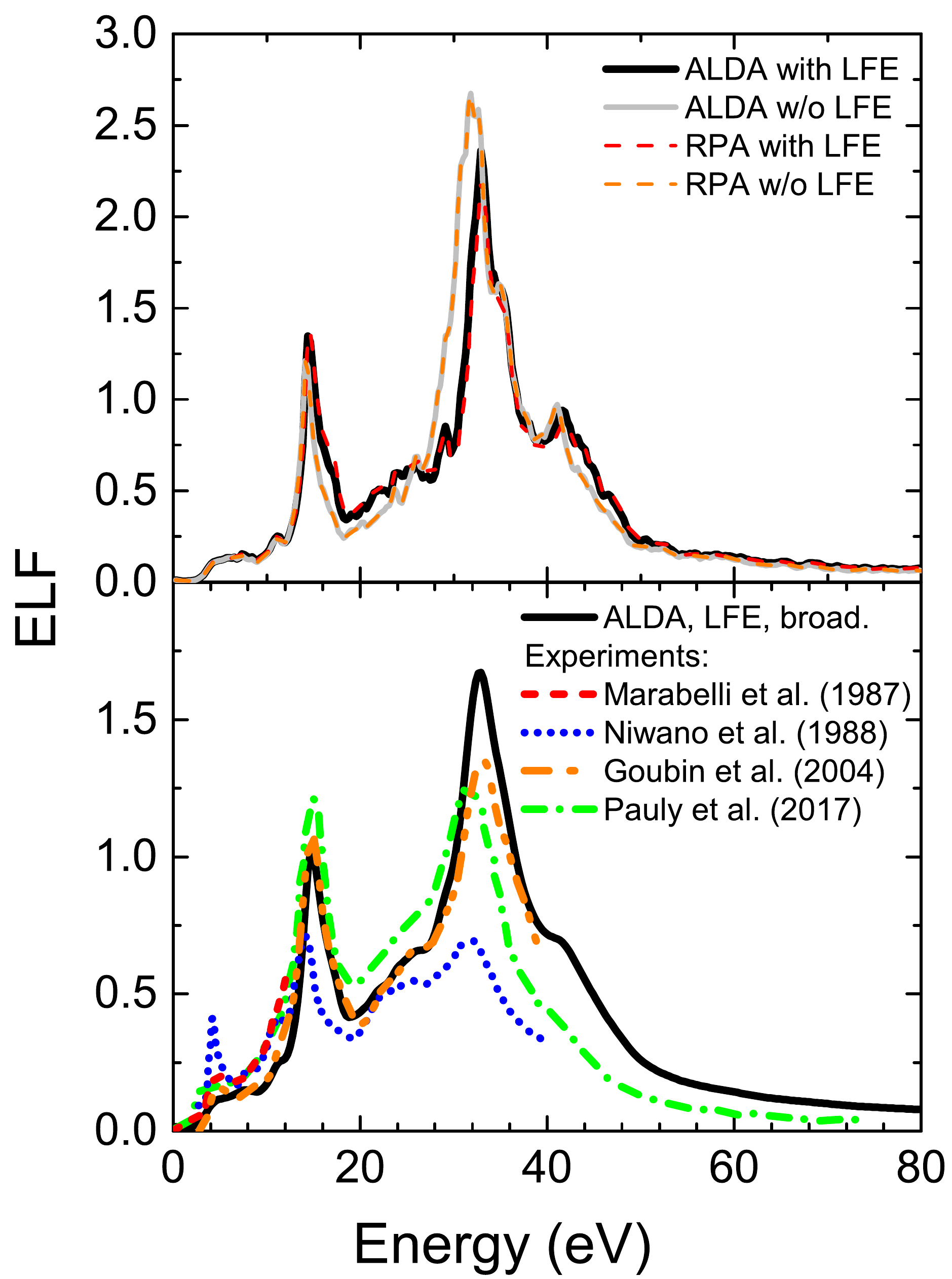}
\caption{Optical energy-loss function of CeO$_2$ under different approximations. Top panel: calculations using ALDA and RPA approximations, with and without LFE. Bottom panel: optical ELF obtained within ALDA and LFE applying peak broadening, in comparison to different sets of experimental data \cite{Marabelli1987,Niwano1988,Goubin2004,Pauly2017}.}
\label{fig:optELFCeO2}
\end{figure}

\begin{figure*}[t]
\centering
\includegraphics[width=0.8\textwidth]{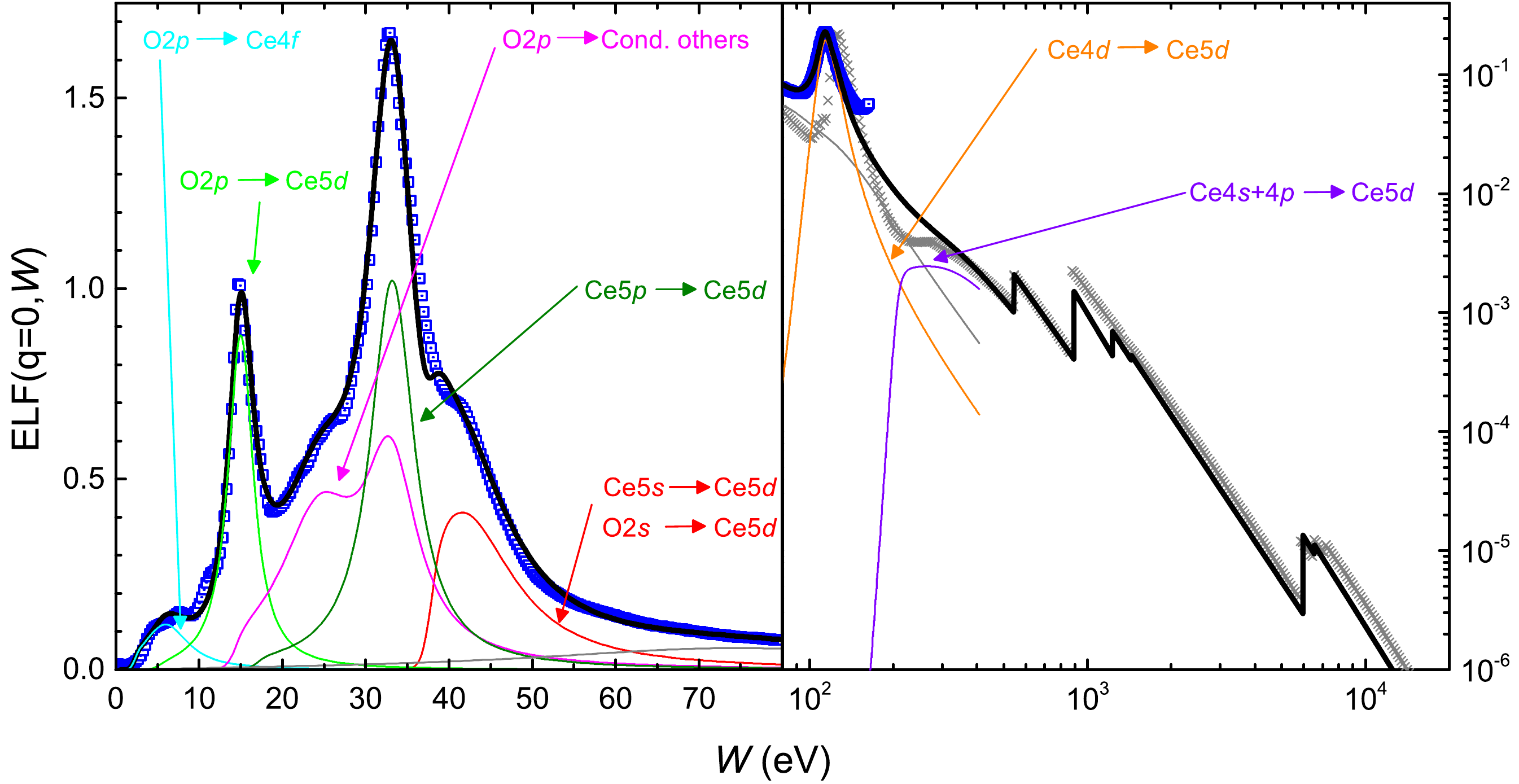}
\caption{Optical ELF of bulk CeO$_2$ at low (left panel) and high (right panel) energies  obtained from ALDA with LFE and considering peak broadening (symbols), together with Mermin ELF fitting for different transitions indicated by labels (solid lines).}
\label{ELFopt1}
\end{figure*}

Fig. \ref{fig:ComparisonEPS1} shows the real ($\bar{\epsilon}_1$, top panel) and imaginary ($\bar{\epsilon}_2$, bottom panel) parts of the macroscopic dielectric function in the optical limit (${\bf{q}}\rightarrow 0$), respectively, in comparison with experimental data \cite{Marabelli1987, Niwano1988, Goubin2004,Pauly2017}. We notice how the inclusion of LFE in the description of the macroscopic dielectric function is critical to reproduce the experimental measurements, in particular in the flattening of the third major peak between 15--20~eV in $\bar{\epsilon}_1$ and between 20--25~eV for $\bar{\epsilon}_2$.\\
\indent The features seen in Fig. \ref{fig:ComparisonEPS1} for $\bar{\epsilon}_2$ can be understood on the basis of the PDOS shown in  the top panel of Fig. \ref{fig:PDOS}. The first peak around 3~eV
corresponds to the transition from the O$2p$ valence band to the Ce$4f$ localized states, with the fundamental band gap of 2.35~eV. 
The second gap extracted from Fig. \ref{fig:PDOS} is of 5.2~eV, above which three main structures can be seen in the $\bar{\epsilon}_2$ spectrum of Fig. \ref{fig:ComparisonEPS1} below 15~eV.
These peaks correspond to the superposition of the PDOS of the O$2p$ occupied levels with that of the Ce$5d$ empty levels.
The next peaks above 15~eV correspond to a mix of transitions from the O$2p$ band to the top
part of the conduction band (being an hybridization of $s$, $p$, $d$ and $f$ states, labelled as ``conduction band (others)'' in Fig. \ref{fig:PDOS}) and from the Ce5$p$ band to the conduction Ce5$d$ band, as previously noted \cite{Khalifi2016}. The former presents an approximate onset of 14.12~eV (a dip in the PDOS of the conduction band should be
noted in Fig. \ref{fig:PDOS} around 14~eV), while the latter has a gap of 16.8~eV (see Fig. \ref{fig:PDOS}), approximately corresponding with the structures observed in Fig. \ref{fig:ComparisonEPS1}. This assignment is in agreement with previous studies \citep{Goubin2004,STRASSER1985765,PhysRevB.30.1155,Pauly2017}. Transitions with $\Delta l=\pm 1$ are allowed, despite angular momentum selection rule, owing to the hybridization between $p, d$ and $f$ orbitals of O and Ce (see top panel of Fig. \ref{fig:PDOS}).
\\
\indent The optical ELF of CeO$_2$ is shown in Fig. \ref{fig:optELFCeO2}. The top panel depicts the TDDFT results using the ALDA kernel or the Random Phase Approximation (RPA), including or not LFE. Both ALDA and RPA give similar results, while LFE play a crucial role in reducing the intensity of the optical ELF, bringing it closer to the experimental data~\cite{Marabelli1987,Niwano1988,Goubin2004,Pauly2017}. The ALDA calculation including LFE is compared to recorded data in the bottom panel, where a broadening to the calculated ELF has been applied in order to account for the finite lifetime of the electronic excitations and the experimental resolution. To take into account the electron-electron and phonon dephasing, we have added an absolute Lorentzian width equal to 0.2~eV, a Lorentzian with quadratic energy dependence multiplied by a factor $6 \times 10^{-4}$ and a Gaussian with $\sigma$=0.2~eV. The calculations agree well with the experimental data below 40~eV, except for intensities of the peaks for Ref.\cite{Niwano1988}.
We notice that the agreement between our calculations and the experimental findings by Goubin et al. \cite{Goubin2004} is remarkable in the entire energy range. Above 40~eV, the calculations overestimate the experiments, although they reproduce very well their shape.
\\
\indent The discrepancy found between our calculations and experiments at the larger energies in Fig. \ref{fig:optELFCeO2} can be rationalised in the light of the $f$- and $ps$-sum rules \cite{Smith1978}, which serve to check the accuracy of an optical ELF. The former gives the effective number of electrons per unit cell participating in the electronic excitations:
\begin{equation}\label{fsumrule}
N_\mathrm{eff}=\frac{2}{\pi\Omega_p^2}\int_0^{W'_{max}}W'\mathrm{Im}\Bigl[\frac{-1}{\bar{\epsilon}(0,W')}\Bigl]dW',
\end{equation}
where $\Omega_p=\sqrt{4\pi {\cal{N}} e^2/m}$, which should converge to the atomic number as $W'_{max} \rightarrow \infty$.
The latter sum-rule states that the ELF should fulfill equation:
\begin{equation}
\frac{2}{\pi}\int_0^\infty \mathrm{Im}\Bigl[\frac{-1}{\bar{\epsilon}(0,W')}\Bigl]\frac{dW'}{W'}+\frac{1}{n_0^2}=1 .    
\end{equation}
The $f$-sum rule fulfillment can be required 
also shell by shell, which allows a more in depth assessment of the optical ELF. \cite{doi:10.1021/acs.jpcc.8b10832} The optical ELF can be fitted by means of the MELF--GOS model, see section \ref{subsec:ELF}. This fitting does not only provide an appropriate extension of the first-principles ELF to large energy and momentum transfers (needed to compute both the sum rules and the electronic cross sections), but can also help to assign the different features of the ELF to particular transitions if the individual Mermin and GOS functions are associated to particular transitions, as suggested in Ref. \cite{doi:10.1021/acs.jpcc.8b10832}. 
\\
\indent Note that, since we are dealing with all-electron simulations, in principle the inclusion of high-energy transitions above 120~eV is possible. However, the extension of the simulations in the energy axis well beyond the value of 120~eV cannot be typically achieved at an affordable computational cost from first-principles calculations.
In this respect, 
this extension of the energy range 
(including 
the Ce and O core levels)
has been performed by means of the atomic GOS within the MELF--GOS model, while at intermediate energies between valence and core excitations, a Mermin function fitting to the atomic X-ray data by Henke et al. \cite{Henke1993} has been carried out. 
\\
\indent The structures observed in the $\bar{\epsilon}_2$ spectrum can be translated to those seen in the optical ELF, where the peaks are generally shifted to larger energies as compared to $\bar{\epsilon}_2$, as ELF=$\mathrm{Im}(-1/\bar{\epsilon})=\bar{\epsilon}_2/(\bar{\epsilon}_1^2+\bar{\epsilon}_2^2)$ (see Figs. \ref{fig:ComparisonEPS1} and \ref{fig:optELFCeO2} for a comparison between energy peaks) and typically $\bar{\epsilon}_1$ decreases with an increase in the energy transfer. Additionally, plasmon excitations can be found by searching for the conditions $\bar{\epsilon}_1\approx 0$ and $\bar{\epsilon}_2\ll 1$. According to our calculations, only the excitations around 12-14~eV and 32~eV could have collective or plasmon-like character.
While several authors \cite{STRASSER1985765,PhysRevB.30.2462,Pauly2017} attribute the first peak to a plasmon excitation, this is not entirely evident from our simulations, even though the inclusion of LFE lowers the minimum of $\bar{\epsilon}_2$ in correspondence of $\bar{\epsilon}_1 \approx 0$ (see Fig. \ref{fig:ComparisonEPS1}), resulting in a maximum of the ELF.
\begin{table}[t]
\caption{\ Assignment of Mermin functions and fulfillment of the individual $f$-sum rule of each interband transition for bulk CeO$_2$.}
\label{tbl:drude1}
  \begin{minipage}{\columnwidth}
  \begin{tabular*}{\columnwidth}{@{\extracolsep{\fill}}lcccc}
    \hline
    Mermin  & Transition & $W_{\rm th}$ (eV)\footnote{Threshold energies derived from the PDOS except otherwise stated} & $N_{\rm exp}$    & $N_{\rm eff}$  \\
    \hline
 1      &   O$2p\rightarrow$Ce$4f$    & 2.35    &       &        \\
2 & O$2p\rightarrow$Ce$5d$ &	5.20 & &	 \\ 	
 & (Plasmon-like)	& & & \\ 	
3-4 & O$2p\rightarrow$Cond. (others) & 14.12 &	&  \\
1-4 & O$2p$,Ce$6s$,$5d$, $4f$\footnote{The Mermin functions 1-4 describe the excitations of the outermost valence electrons, located in a band of mainly O$2p$ origin, but which also contains the valence electrons Ce$6s$, $5d$, and $4f$, see Fig. \ref{fig:PDOS}} &  &	12$^b$	& 8.08 \\
5 &	Ce$5p\rightarrow 5d$ & 16.80	& 6	 & 6.13 \\
 & (Plasmon-like) &	& & \\ 	
6 &	Ce$5s$,O$2s\rightarrow$Ce$5d$ & 37.80\footnote{From Ref. \cite{Williams}} &	6 &	5.91 \\
7 &	Not assigned & & & 5.32 \\
8-9 &	Ce$4d\rightarrow 5d$ & 100.68$^c$ &	10 &	10.87\\
10 &	Ce$4s,4p\rightarrow 5d$ & 206.53$^c$ &	8 &	7.89 \\
11 &	Not assigned & & & 6.39 \\ \hline
GOS & Atomic shells &	& \\ \hline
O$K$      &   O$1s$ & 543.10$^c$   &    4       &    3.48 \\
Ce$M$ & Ce$3s,3p,3d$ & 883.80$^c$ & 18 & 14.23	\\
Ce$L$ & Ce$2s,2p,4f$ & 5723.00$^c$ & 8	& 4.53 \\
Ce$K$ & Ce$1s$ & 40443.00$^c$	&  2 &	1.07 \\ 
\hline
Total &	 & &	74 &	73.89  \\
\hline
  \end{tabular*}
\end{minipage}
\end{table}
\\
\indent Following the trends in the band gaps and $\bar{\epsilon}_2$ features obtained from the PDOS analysis, we fitted the first-principles optical ELF of CeO$_2$ with 11 Mermin ELFs assigned to particular transitions, see Fig. \ref{ELFopt1}. However, two of the Mermin functions (7 and 11, shown by a gray line) were intended to act as ``baseline'' to obtain a smooth crossover between valence and core levels and, thus, have not been assigned to any particular transition  to fulfill the sum rules. Note that for the larger energies (Ce$5s$, O$2s$ and larger energy excitations, including those described by GOS), 
threshold energies have been taken from atomic binding energies from the X-Ray Data Booklet \cite{Williams}, see Table \ref{tbl:drude1}. 
The individual contributions to the Mermin function fitting of the first-principles optical ELF are depicted by thin solid lines in Fig. \ref{ELFopt1} and the corresponding parameters are given in Table S1 in the Supplementary Information. Each Mermin function or group of functions has been assigned to particular transitions, as indicated in Table \ref{tbl:drude1}. The assignment has been made  with the purpose of fulfilling the individual $f$-sum rules, i.e. that the effective number of electrons $N_{\rm eff}$ coming from each interband transition is close to the number of electrons expected in such transition $N_{\rm exp}$.\\
\indent 
To this respect, two further aspects should be taken into account. First, the Mermin functions 1 to 4 are associated with transitions from the occupied O$2p$ band, as this is the main hybridization of the valence band. However, this being the outermost band, it also hosts the valence electrons from Ce$6s,5d,4f$, so the total expected number of electrons there should be 12. Second, the inner shells, described by hydrogenic GOS, systematically underestimate the number of inner-shell electrons (see Table \ref{tbl:drude1}). In CeO$_2$ the number of core electrons in the O$K$, Ce$K,L,M$ shells is 32, while we find $\approx 23$. Thus, the remaining electrons are relocated in the outer shells in order to fulfil the total $f$-sum rule. These missing electrons are located in the unassigned ``baseline'' Mermin functions (7 and 11), so they do not affect much the electron count in the other assigned contributions.
\\
\indent Despite some underestimation in the number of electrons in a few shells (e.g. of the 12 O$2p$ and Ce$4f$, $5d$ and $6s$ electrons, only 8.08 electrons were found, see  Table \ref{tbl:drude1}), the assignment of the levels is rather satisfactory for most of the transitions. 
The $\sim 4$ electrons missing from the O$2p$ and Ce$6s,5d,4f$ bands (Mermin functions 1 to 4) are contained in the seventh not assigned Mermin function (see Table \ref{tbl:drude1}), which possibly may represent a large energy tail of these transitions.
The fulfilment of the total $f$- and $ps$-sum rules is remarkable, with errors of $-0.15$\% and $+0.36$\%, respectively. 
Particularly, the effective number of electrons in our fitting corresponding to the Ce$5p\rightarrow5d$ (dark green line in Fig. \ref{eps1_eps2_ELF}) and Ce$5s$,O$2s\rightarrow$Ce$5d$ transitions (red line) is very close to the expected number of electrons. This reveals an underestimation in the $f$-sum rule of the experimental ELFs\cite{Marabelli1987,Niwano1988,Goubin2004,Pauly2017} in the range 30--70~eV, which may reveal some degree of experimental uncertainty. The Mermin functions corresponding to the Ce$4d\rightarrow 5d$ (orange line) and Ce$4s,4p\rightarrow 5d$ transitions (purple line) also deliver effective number of electrons remarkably close to the expected ones.\\
\indent Finally, to determine the inelastic scattering cross sections within the dielectric approach \cite{Ritchie1957,Ritchie1959,Dapor2020book}, we need to assess the dependence of the ELF in both the energy and  momentum space (the so-called Bethe surface). 
We report in Fig. S6 of the Supplementary Information the energy and momentum dependent ELF of bulk CeO$_2$ calculated according to eqns (\ref{eq:eM}) and (\ref{ELF}) in the range $0-6$~\AA$^{-1}$ along the [111] direction in reciprocal-space for energies up to $120$~eV. We notice that the agreement between MELF--GOS and TDDFT results is reasonable at low energy and finite momentum, while the peak at 120~eV shows a momentum dispersion that does not increase in energy (see Fig. S6 of the Supplementary Information).
We assessed the dependence on the direction of the momentum transfer of the ELF of bulk CeO$_2$, finding a small difference between the [111], [110], [211] orientations (see Fig. S7 of the SI showing this dependence). Therefore, we used the ELF in the [111] direction for the Monte Carlo simulations of the REEL spectrum.

\subsection{The excitation spectrum of bulk Ce$_2$O$_3$}
\label{sec:Ce2O3}

\begin{figure}[t]
\centering
\includegraphics[width=\columnwidth]{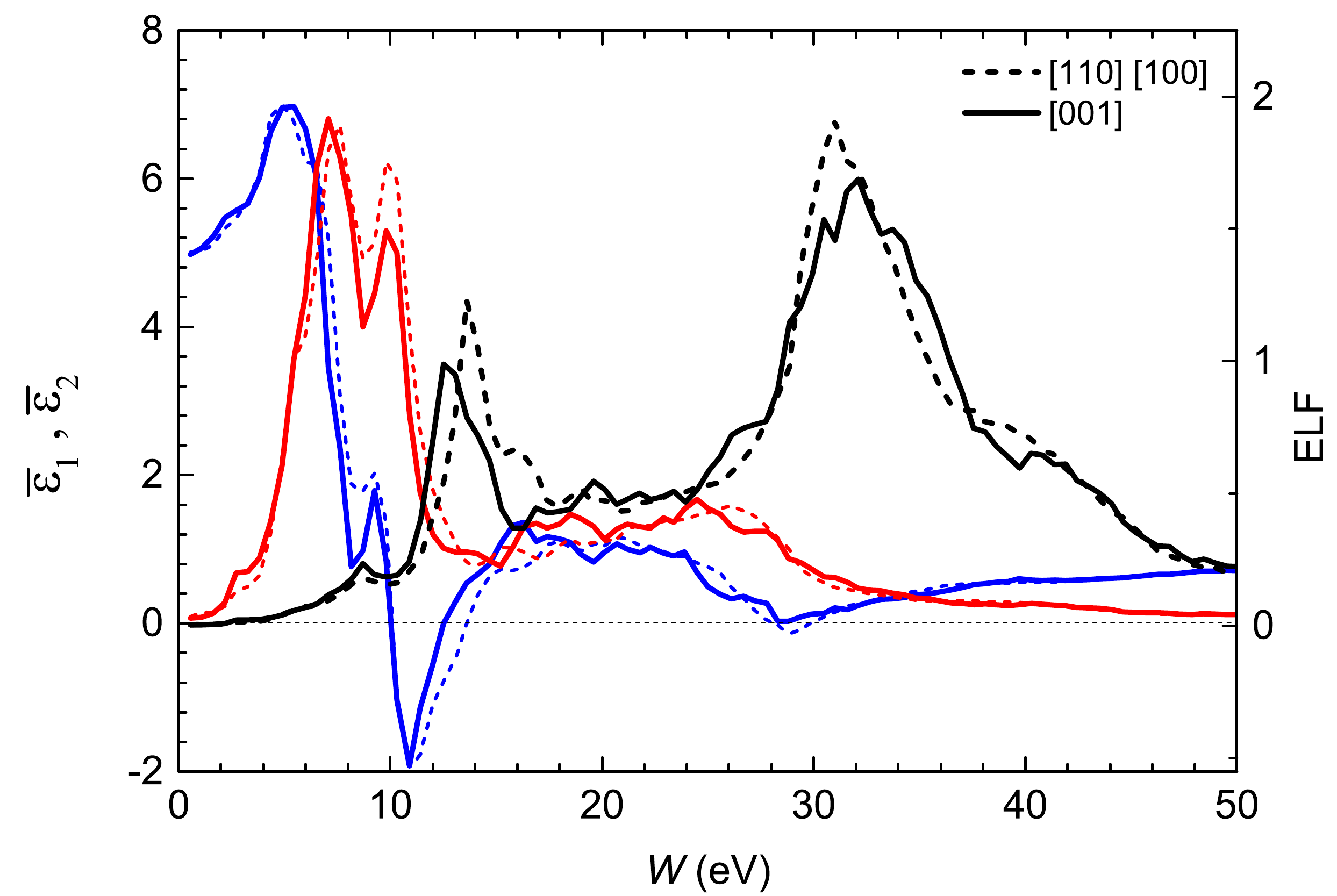}
\caption{Real ($\bar{\epsilon}_1$, blue lines) and imaginary ($\bar{\epsilon}_2$, red lines) part of the macroscopic dielectric function of bulk Ce$_2$O$_3$ along with the optical ELFs. $\bar{\epsilon}_1$, $\bar{\epsilon}_2$ and the optical ELF are presented along the [001] (solid lines) and the [110] and [100] directions (dashed lines) of the momentum transfer.}
\label{eps1_eps2_ELF}
\end{figure}

\indent To identify the electronic transitions of bulk Ce$_2$O$_3$ we followed an approach similar to that of CeO$_2$, based on the MELF--GOS fit. The first-principles real ($\bar{\epsilon}_1$) and imaginary ($\bar{\epsilon}_2$) parts of the macroscopic dielectric function, together with the resulting ELF in the [001] (solid lines) and the equivalent [110] and [100] directions (dashed lines) are shown in Fig. \ref{eps1_eps2_ELF}, taking the ${\bf{q}} \rightarrow 0$ limit. Although there is a small difference in the crystal direction, it will be seen that it is not as significant to affect the REEL Monte Carlo simulations.
\\
\indent From the combined analysis of the PDOS (see Fig. \ref{fig:DOSCe2}) and of Fig. \ref{eps1_eps2_ELF} we can argue that the first shoulder around 5~eV corresponds to the Ce4$f\rightarrow$Ce$4f,5d$ transition, with a gap of 2.04~eV. The next two peaks correspond to transitions from the O$2p$
band to the narrow Ce$4f$ band, and from the O$2p$ band to the wider Ce$5d$ band, with
threshold 3.83~eV. The first peak above 13~eV corresponds to another O$2p$ transition
to conduction band, composed of a mixture of states, with
approximated threshold 13.06~eV (notice the dip in the TDOS at this energy in Fig. \ref{fig:DOSCe2}). The
other peaks above 13~eV can be attributed to transitions from the Ce$5p$ band to the Ce$4f,5d$ band, with threshold of 15.10~eV. The O2$s\rightarrow$ Ce$4f,5d$ transitions can be found at higher energies.\\
\begin{figure*}[t]
\centering
\includegraphics[width=0.8\textwidth]{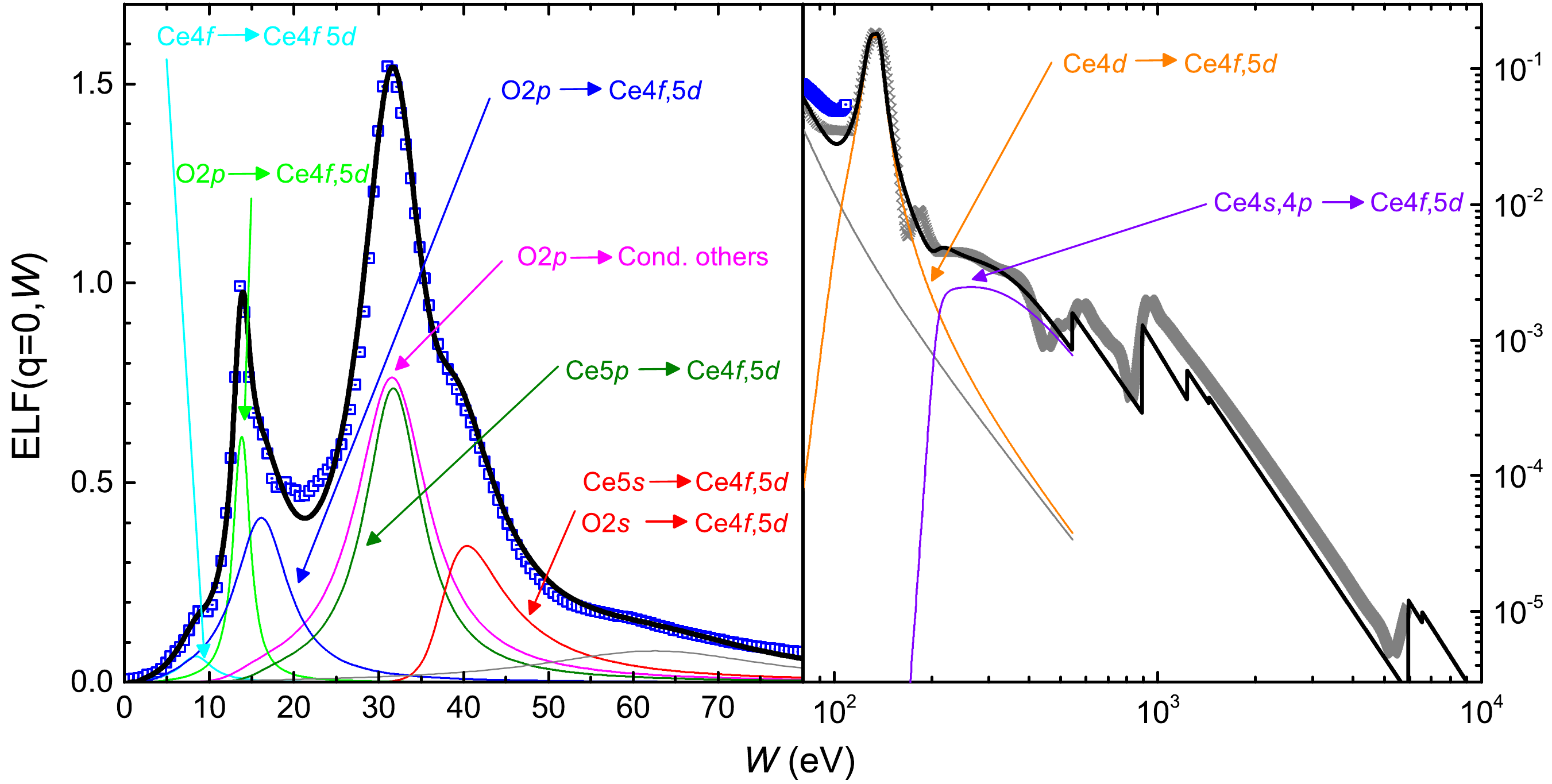}
\caption{Optical ELF of bulk Ce$_2$O$_3$ at low (left panel) and high (right panel) energies obtained from ALDA with LFE and considering peak broadening (symbols), together with Mermin ELF fitting for different transitions indicated by labels (solid lines).}
\label{ELFopt2}
\end{figure*}
\indent The optical ELF of bulk Ce$_2$O$_3$ is fitted using the MELF--GOS model 
following the above assignments and band gaps, and we report its low ($< 70$~eV) and high (up to 10~keV) energy structure in Fig. \ref{ELFopt2}. As seen from Table \ref{tbl:drude2}, a number of 17.12 electrons are missing from the inner shells, which are relocated among the outer shells.
The first Mermin function (cyan curve) represents the Ce$4f$ (occupied) $\rightarrow$ Ce$4f,5d$ (empty) transition.
The transitions from the O$2p$ band (also containing the outermost Ce$4f,5d,4s$ electrons) can be separated in three distinguishable contributions. The green line corresponds to the transition from the O$2p$ levels to the narrow empty band of Ce$4f$ character with threshold 3.83~eV, and possibly having a plasmon-like character, as suggested by the results of Fig. \ref{eps1_eps2_ELF}. The blue line represents the
transition to the wider Ce$5d$ band with threshold 3.83~eV, and finally the magenta line identifies the
transition to the upper part of the conduction band  with an approximate threshold of 13.06~eV. A test of the $f$-sum rule 
by integrating all the 1-4 Mermin functions delivers 20.68 electrons, comparable to the expected value of 20. 
\\
\indent As in the case of CeO$_2$, the Ce$5s$,O$2s \rightarrow$ Ce$4f,5d$ transition (red line) is located at higher energies
with a threshold of 37.8~eV (from Ref.\cite{Williams}). The $f$-sum rule test delivers 10.10 electrons of the 10 electrons expected. 
A seventh Mermin function (gray line) has not been assigned, and recovers a major part (11.81 of 17.12) of the electrons missing from the inner shells. 
Finally, we identify the transitions Ce$4d\rightarrow$Ce$5d$ lying above 10.6~eV (orange line, 24.47 electrons of the 20 expected, containing also some of the electrons missing from the inner shells) and Ce$4s,4p \rightarrow$Ce$5d$ (purple line, 15.98 electrons of the 16 expected).
We notice 
the satisfactory fulfillment of the total $f$-sum rule with an error of 0.14\%, and of the $ps$-sum rule with an error of 0.70\%.
\\
\indent The energy and momentum dependent ELF of bulk Ce$_2$O$_3$ along the [110] direction in reciprocal-space from 0 to 6 \AA$^{-1}$ is provided in the bottom panel of Fig. S6 of the Supplementary Information. As for CeO$_2$, we also checked the dependence on the direction of the momentum transfer of the ELF of bulk Ce$_2$O$_3$, finding a difference between the [001], [110], [100] orientations (depicted in Fig. S8 of the Supplementary Information) without significant effects on MC simulations of REEL spectra. 
Thus, we used the ELF structure in the specific direction [110] for the momentum transfer in MC simulations.
\\
\begin{table}[h!]
\caption{\ Assignment of Mermin functions and fulfillment of the individual $f$-sum rule of each interband transition for bulk Ce$_2$O$_3$.}
\label{tbl:drude2}
\begin{minipage}{\columnwidth}
  \begin{tabular*}{\columnwidth}{@{\extracolsep{\fill}}lcccc}
    \hline
    Mermin & Transition &  $W_{\rm th}$ (eV)\footnote{Threshold energies derived from the PDOS except otherwise stated} & $N_{\rm exp}$    & $N_{\rm eff}$   \\
    \hline
 1      &   Ce$4f\rightarrow4f,5d$    & 2.04    &  &                  \\
2 & O$2p \rightarrow$Ce$4f,5d$	& 3.83 & & \\
& (Plasmon-like?) & & & \\
3 & O$2p\rightarrow$Ce$4f,5d$ & 3.83 & & \\
4 &	O$2p\rightarrow$ Cond. (other) & 13.06	&	 & 	 \\
1-4 & O$2p$, Ce$6s$, Ce$5d$, Ce$4f$\footnote{The Mermin functions 1-4 describe the excitations of the outermost valence electrons, located in a band of mainly O$2p$ origin, but which also contains the valence electrons Ce$6s$, $5d$, and $4f$, see Fig. \ref{fig:PDOS}} &  & 20$^b$ & 20.68 \\
5 &	Ce$5p\rightarrow 4f,5d$ & 15.10 & 12 &	12.16 \\
& (Plasmon-like?) & & & \\
6 &	Ce$5s$,O$2s\rightarrow$Ce$ 4f,5d$ & 37.8\footnote{From Ref. \cite{Williams}} &	10 &	10.10 \\
7 &	Not assigned & &  & 11.81	 \\
8-9 &	Ce$4d\rightarrow 4f,5d$ & 100.68$^c$ &	20 & 24.47  \\
10 &	Ce$4s,4p\rightarrow4f,5d$ & 206.53$^c$ & 16 & 15.98 \\
\hline
GOS  & Atomic shells &     &   &  \\
   \hline
O$K$ &  O$1s$ & 543.68$^c$  &    6   &    5.22               \\
Ce$M$ & Ce$3s,3p,3d$ & 883.80$^c$ & 36 & 28.47	 \\ 
Ce$L$ & Ce$2s,2p$ & 5723.00$^c$ & 16	& 9.06 \\
Ce$K$ & Ce$1s$ & 40443.00$^c$	&  4 &	2.13  \\ 
\hline
Total & & &	140 & 140.19 \\
\hline
  \end{tabular*}
\end{minipage}
\end{table}

\subsection{Electron transport in the cerium oxides}
\label{sec:REELSMC}

\begin{figure}[t]
\centering
\includegraphics[width=\columnwidth]{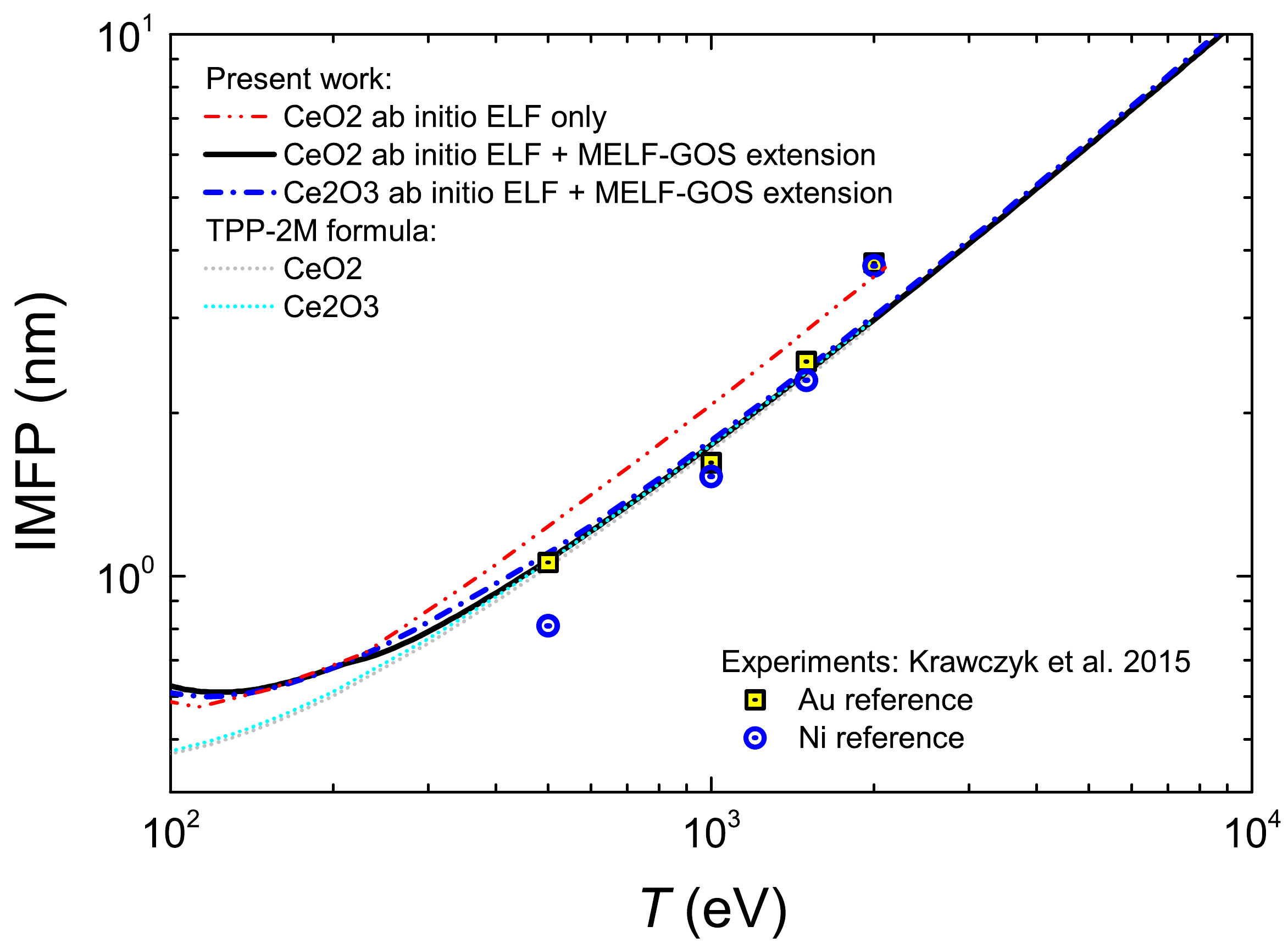}
\caption{Comparison of the ab initio IMFP of bulk CeO$_2$ and Ce$_2$O$_3$ 
calculated in this work 
with TPP--2M formula predictions 
and experimental data for CeO$_2$ obtained with respect to Ni and Au substrates \citep{Jablonski2015}.}
\label{fig:DIIMFP}
\end{figure}

The description of the inelastic interactions of charged particles moving through bulk solids in terms of the DIIMFP (see eqn (\ref{EqDIIMFP})) and of the IMFP (see eqn (\ref{EqIMFP})) is one of the input information to carry out charge transport MC simulations. We thus calculated the latter observables for bulk CeO$_2$. We plot the IMFP in 
Fig. \ref{fig:DIIMFP}, alongside the experimental data obtained on Ni (circles) and Au (squares) substrates \citep{Jablonski2015}. 
While the agreement between calculations and experimental measurements is remarkable, we stress the crucial importance of including the semicore $4d\rightarrow 5d$ transitions and deeper inner-shell excitations (those beyond $100$~eV) in the assessment of the IMFP. Indeed, using only the ab initio data limited to 100~eV (the typical energy range affordable for TDDFT calculations) we notice a discrepancy up to 25\% higher with respect to the experimental data (see the difference between the orange dot-dot-dashed and solid black curves around 120~eV in Fig. \ref{fig:DIIMFP}.) 
\\
\indent The 
IMFP of bulk Ce$_2$O$_3$ is also reported
as a blue dot-dashed line in Fig. \ref{fig:DIIMFP}, showing no significant differences with that of CeO$_2$. It should be noted that, despite of the resemblance of the IMFP of both oxides, they present some differences (not shown here explicitly) in their DIIMFP, eqn (\ref{EqDIIMFP}), arising from their diverse electronic excitation spectra (see Figs. \ref{ELFopt1}, \ref{ELFopt2} and S6).
\\
\indent Typically, elastic scattering occurs between electrons and the massive ionic constituents of the target, and results only in the deflection of electron trajectories. While in principle relativistic first-principles quantum mechanical calculations based on the formal theory of scattering \cite{taioli2015computational,taioli2010electron} can be used to calculate the DESCS \cite{taioli2020relative},  
however the computational cost to include the large number of atoms required to carry out accurate relativistic simulations in a solid is prohibitive. Thus, we use the expression of the Mott's cross section (see eqns (\ref{scatel}) and (\ref{molsca})) within the cluster method, by which
we carve out a portion of the material (which can be as small as one atom) and reckon the DESCS only dealing with a molecular cluster surrounded by empty space.
This approximation, which can be made more accurate by increasing the cluster size until the DESCS ceases to vary, can be drastic at times but works in our case.
However, the exclusion (inclusion) of multiple scattering effects in the calculation of the DESCS reckoned by using eqn (\ref{scatel}) (\ref{molsca}) leads to negligible discrepancies. 
Thus, using eqn (\ref{scatel}) we have calculated the DESCSs of CeO$_2$ and Ce$_2$O$_3$ at different primary beam kinetic energies.
In DESCS calculations the electrostatic potential in the Dirac equation was modelled by a screened Coulomb interaction. The latter is obtained by multiplying a bare Coulomb potential by a function expressed as a superposition of Yukawa functions, whose parameters were set according to a best fit of data from Hartree-Fock simulations \cite{doi:10.1063/1.1595653}. Exchange effects were described by using the Furness and McCarthy formula \cite{Furness_1973}.

\begin{figure}[t]
\centering
\includegraphics[width=\columnwidth]{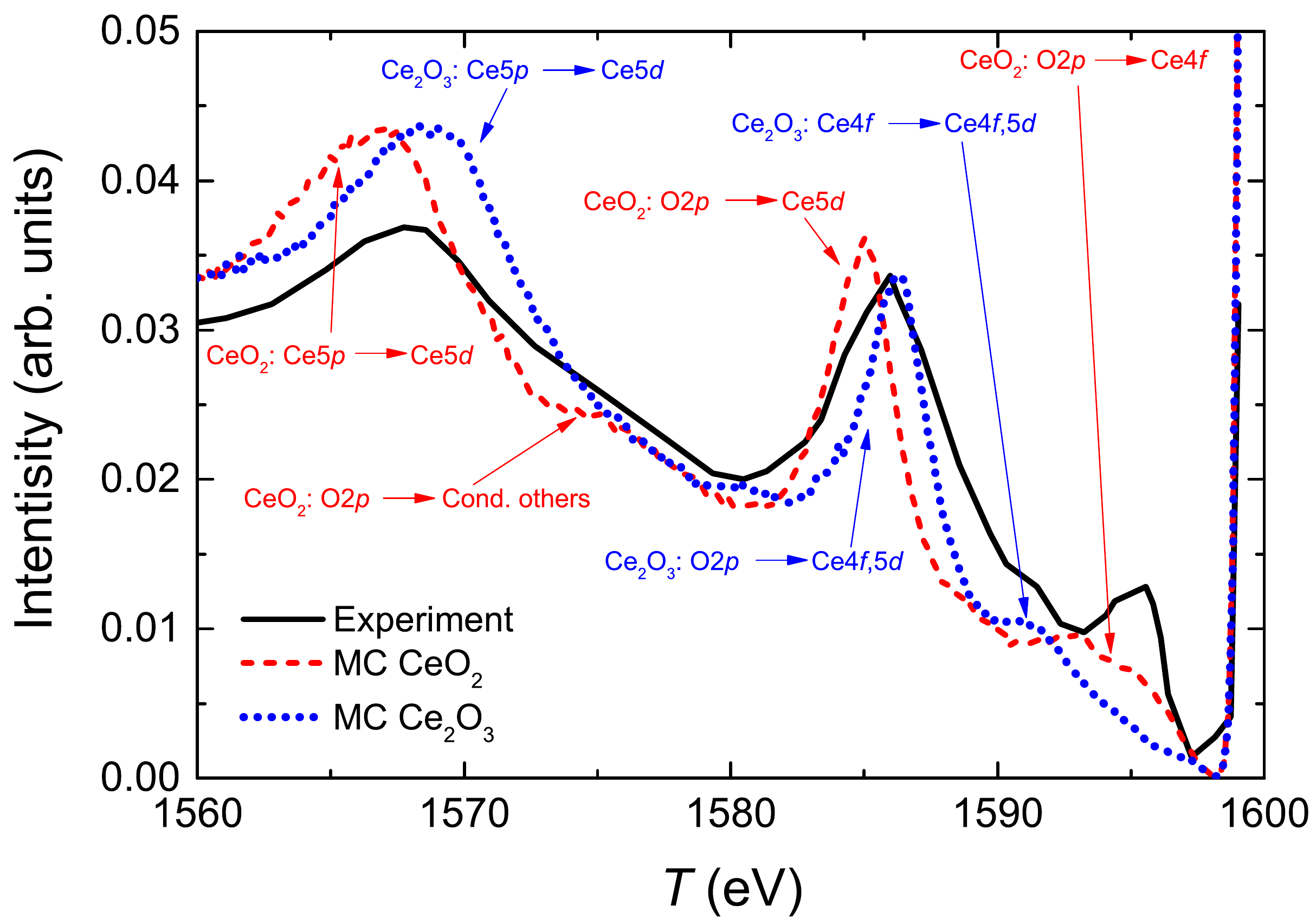}
\caption{Comparison between the REEL spectra of bulk CeO$_2$ (red line) and Ce$_2$O$_3$ (blue line) simulated by MC compared with the directly measured spectrum (black line) \citep{Pauly2017}. Spectra are normalized with respect to the area under the curves. Zero-loss peak is found at the extreme right of the spectrum. Primary beam impinges the material with an incident angle equal to 60$^{\circ}$ with respect to the surface's normal.}
\label{fig:REELS-CeO2}
\end{figure}

The REEL spectra of the materials under investigation have been computed by our MC approach, where the electron trajectories ensemble is set to reach statistical significance and low noise of the simulated data ($\approx 10^9$ trajectories). We remind that the REEL spectrum probes the dielectric response of materials in the longitudinal direction (along the momentum transfer), being a purely longitudinal field a condition of propagation of the plasmon. Furthermore, the microscopic components of the transverse induced field are negligible in the optical limit.\\
\indent In Fig. \ref{fig:REELS-CeO2} we report the REEL spectra of bulk CeO$_2$ (red line) and Ce$_2$O$_3$ (blue line), and we also compare our simulations with the available experimental data \cite{Pauly2017}, finding a good agreement by reproducing all the main features of the experimental spectrum. 
These simulated REEL spectra have been obtained using the ELF of CeO$_2$ and Ce$_2$O$_3$ along the [111] and [110] directions in the reciprocal space, respectively.
However, we noticed a negligible dependence on the momentum transfer direction of the REEL spectra (see Figs. S7 and S8 of the SI, where we compare 
the calculated Bethe surfaces along
different orientations of the momentum transfer). The REEL spectra are both normalized with respect to the area under the curves. We remind that surface effects, such as the presence of surface plasmons, are not included in our bulk simulations, while they may affect the measurements. This can explain e.g. the discrepancy of the brighter experimental peak intensity near the elastic peak with our simulations. \\
\indent The overall satisfactory agreement with the experimental data makes us confident that both the simulated dispersion law with the momentum transfer and the MELF--GOS extension to very high energy are indeed accurate. We stress again the need of including the Ce$4d$ transitions and LFE in the calculation of the dielectric response to achieve a good accuracy of the REEL spectra. We notice that the experimental spectrum \cite{Pauly2017} shows features reproduced in both our simulated REEL spectra of CeO$_2$ and Ce$_2$O$_3$ bulk structures, thus experimental determinations seem more consistent with polycrystalline samples possibly mixing these two stable allotropic forms of cerium oxide. \\
\indent The REEL spectra are characterised by a number of structured peaks, due to plasmon-like losses and single electron intraband and interband transitions. In particular, the REEL spectral lineshape of CeO$_2$ in Fig. \ref{fig:REELS-CeO2} (red line) shows three major structures below the elastic zero-energy loss peak of backscattered electrons (extreme right of the plot). According to our previous analysis of both the PDOS and $\bar{\epsilon}_2$, the first peak found at $\approx$ 3.7~eV from the elastic peak can be attributed to the interband transition from the O$2p$ valence band to the localized Ce$4f$ narrow band, the second peak around 15~eV to a plasmon-like excitation from O$2p \rightarrow$ Ce$5d$ reflecting the PDOS of Ce$5d$ conduction band, and finally the third broader loss structure at 32.5~eV from the elastic peak can be attributed to the Ce$5p\rightarrow 5d$ interband transition.\\
\indent Analogously, the REEL spectrum of bulk Ce$_2$O$_3$ reflects the previous analysis of the 
peaks found in the PDOS and in the ELF, showing a lineshape with orbital characters similar to CeO$_2$. Interestingly, the experimental REELS \cite{Pauly2017} shows three structures (a first peak around 1595~eV, a shoulder around 1598--1590~eV, and a second peak at 1575~eV), which are revealed to distinctly correspond to CeO$_2$ (red dashed line) and Ce$_2$O$_3$ (blue dotted line), respectively. Thus, the current approach demonstrates how first principle-based simulations may help to characterize and identify the features observed in experimental determinations.
Finally, we notice that the REEL spectra are almost independent of the chosen direction of the momentum transfer. 

\section{Conclusions}

In this work, we have computed the dielectric response functions of cerium oxides, which are materials with potential applications ranging from radiotherapy to catalysis industry, using linear response TDDFT on the top of ground state calculations based on DFT+U, adding the Hubbard correction to deal with the presence of $4f$ localized states. We found excellent agreement with most of the experimental determinations, showing in particular that an accurate assessment of the optical properties must include LFE. \\
\indent We performed also the assignment of the different features of the spectra in terms of interband transitions, which allowed to further check the accuracy of the results by means of sum rules.
The information collected via numerical simulations based on first-principles was extended to higher energies by the MELF--GOS model to include core-level excitations. We stress that the inclusion of the interband transitions from the $4d$ levels of Ce and of the innermost Ce and O shells, as well as of the LFE into the dielectric response of these oxides is of paramount importance for calculating accurately the IMFP of electrons in their motion through the medium. \\ 
\indent Furthermore, the ELF calculations of both bulk CeO$_2$ and Ce$_2$O$_3$ have also been performed by considering the full energy-momentum dispersion at finite values of momentum transfer. The inelastic collision parameters, directly obtainable from the ELF, along with the elastic scattering ones, are essential ingredients to feed a MC routine and follow the trajectory of electrons within the solid. In this respect, we were able to reproduce (or predict) in remarkably good agreement with the available experimental data the major REEL spectral features of bulk CeO$_2$ and Ce$_2$O$_3$.\\
\indent Our work, based on a statistical MC method informed by high-accuracy ab initio inputs, paves the way towards a thorough understanding of the electron transport and optical properties of cerium oxides, which can drive a proper design of materials, also in nanoparticle shape, with potential disruptive impact on enhancing local radiation damage in cancer cure treatments and catalysis applications.

\section*{Acknowledgements}

The authors acknowledge Bruno Kessler Foundation (FBK) for providing unlimited access to the KORE computing facility.
A.P. acknowledges Fondazione Caritro and DICAM (University of Trento) for the financial support under the CARITRO project {\it High-Z ceramic oxide nanosystems for mediated proton cancer therapy}. A.P. also acknowledges fruitful discussion with the ELK code developers. 
This project has received funding from the European Union's
Horizon 2020 Research and Innovation programme under the Marie Sklodowska-Curie grant agreement N. 840752 {\it NanoEnHanCement} of P.d.V., as well as from the Spanish Ministerio de Econom\'{\i}a y Competitividad and the European Regional Development Fund (Project no. PGC2018- 096788-B-I00), from the Fundaci{\'o}n S{\'e}neca (Project no. 19907/GERM/15) and from the Conselleria d’Educaci{\'o}, Investigaci{\'o}, Cultura i Esport de la Generalitat Valenciana (Project no. AICO/ 2019/070).

\section*{Conflicts of interest}

There are no conflicts to declare.


\balance


\bibliography{Biblio} 
\bibliographystyle{rsc} 

\newpage
\setcounter{section}{0}
\setcounter{figure}{0}
\setcounter{table}{0}
\renewcommand\thefigure{S\arabic{figure}}

\renewcommand{\thetable}{S\arabic{table}}

\onecolumn
\begin{center}
\Huge SUPPORTING INFORMATION

\end{center}
\vspace{1cm}
\section{DOS: tests of different $U_\mathrm{eff}$ values. Band structure of bulk CeO$_2$ and Ce$_2$O$_3$ }

\begin{figure}[htbp!]
\centering
\includegraphics[width=0.7\linewidth]{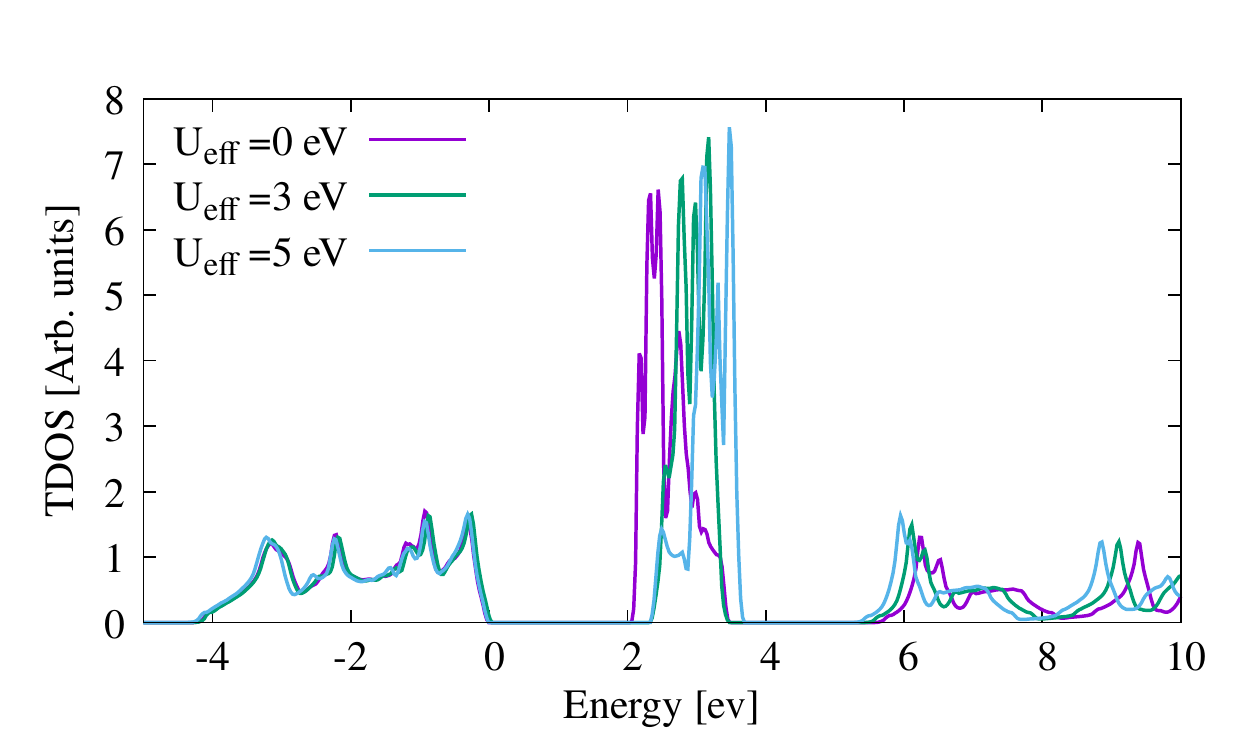}
\caption{Total DOS of bulk CeO$_2$. We show the effect of different Hubbard corrections ($U_\mathrm{eff}=0,3,5$ eV) to deal with the localized $4f$ states (highest peak between 2 and 4 eV). The top of the valence band is set to 0 eV.}
\label{fig:DOS1}
\end{figure}

Figure \ref{fig:DOS1} shows the total density of states (TDOS) of bulk CeO$_2$ obtained by tuning the Hubbard correction in the commonly used range $U_\mathrm{eff}=[0:5]$ eV. \citep{Loschen2007}
Increasing $U_\mathrm{eff}$ results in a blueshift of the $4f$ states localized in the energy region 2 to 4 eV (the top of the valence band is shifted to 0 eV), which further opens the O$2p$-Ce$4f$ gap by $0.3$ eV with respect to the plain DFT-LSDA value. The broad band between $-4$ eV and 0 eV remains untouched by changing $U_\mathrm{eff}$. 

\begin{figure}[htbp!]
\centering
\includegraphics[width=0.7\textwidth]{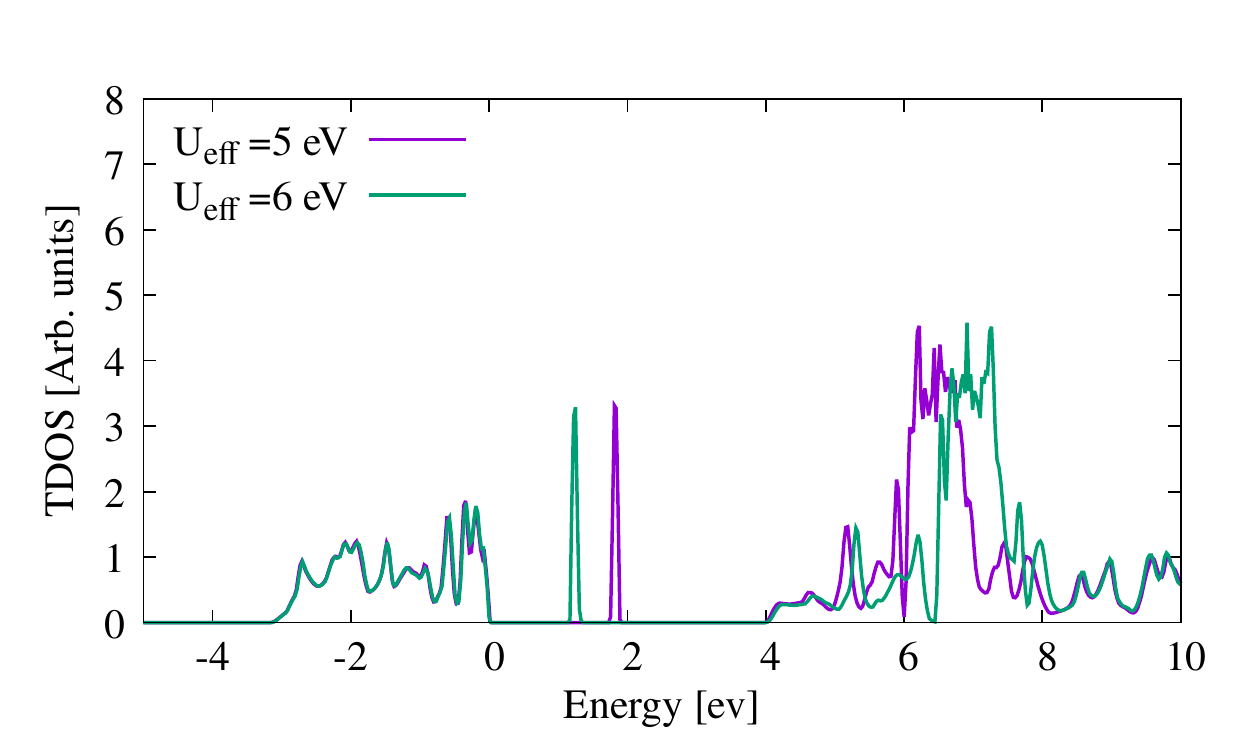}
\caption{
Total DOS of Ce$_2$O$_3$ bulk solid. We show the effect of different Hubbard corrections $U_\mathrm{eff}=5,6$ eV to deal with the localized $4f$ states (sharp peak near to $2$ eV).}
\label{fig:DOS2}
\end{figure}
\begin{figure}[htbp!]
\centering
\includegraphics[width=0.7\textwidth]{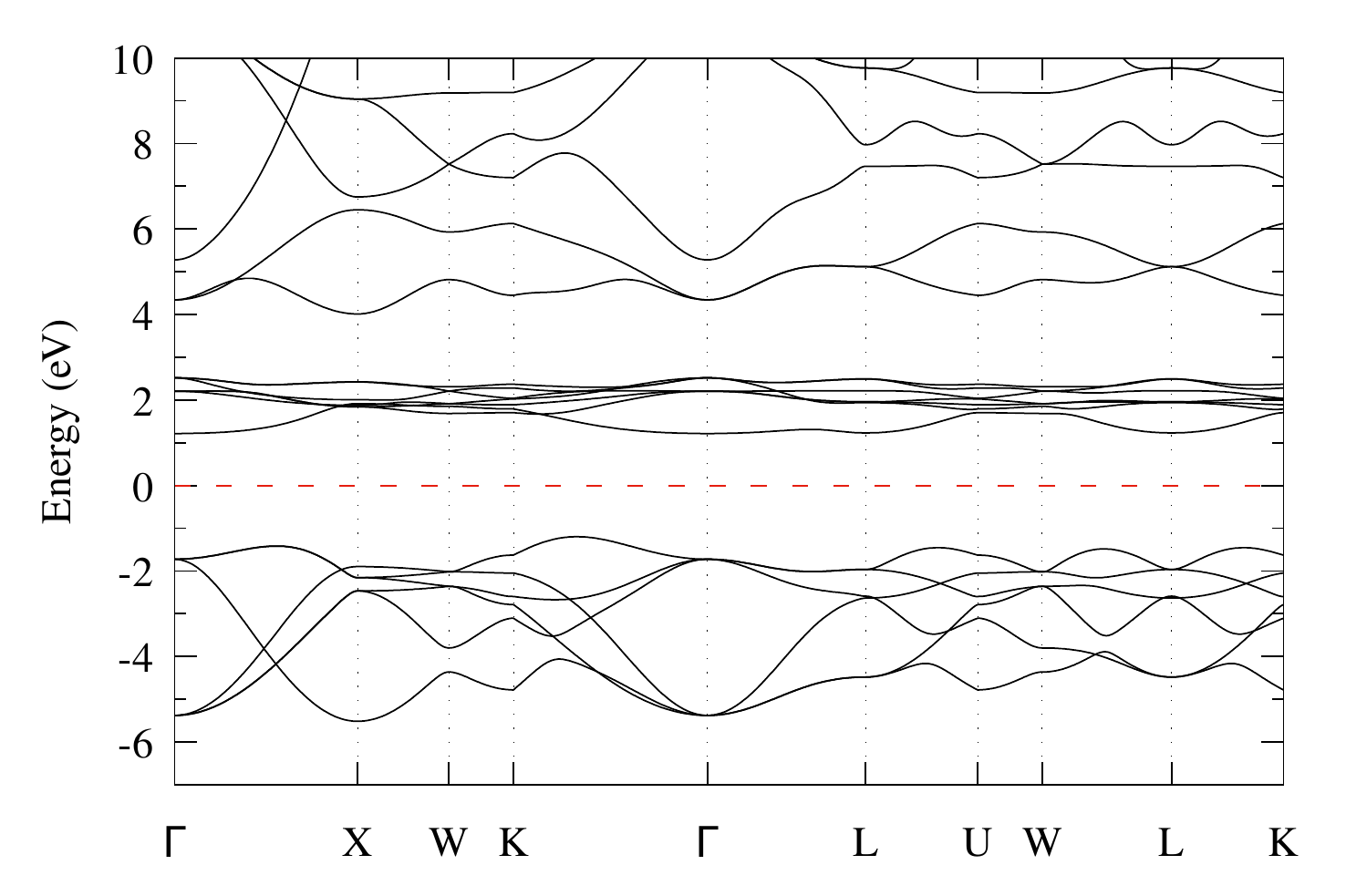}
\caption{Band structure of bulk CeO$_2$ calculated using DFT+U. The Fermi level (dashed red line) is set to the origin of the energy axis.}
\label{Band_CeO2}
\end{figure}
\begin{figure}[htbp!]
\centering
\includegraphics[width=0.7\textwidth]{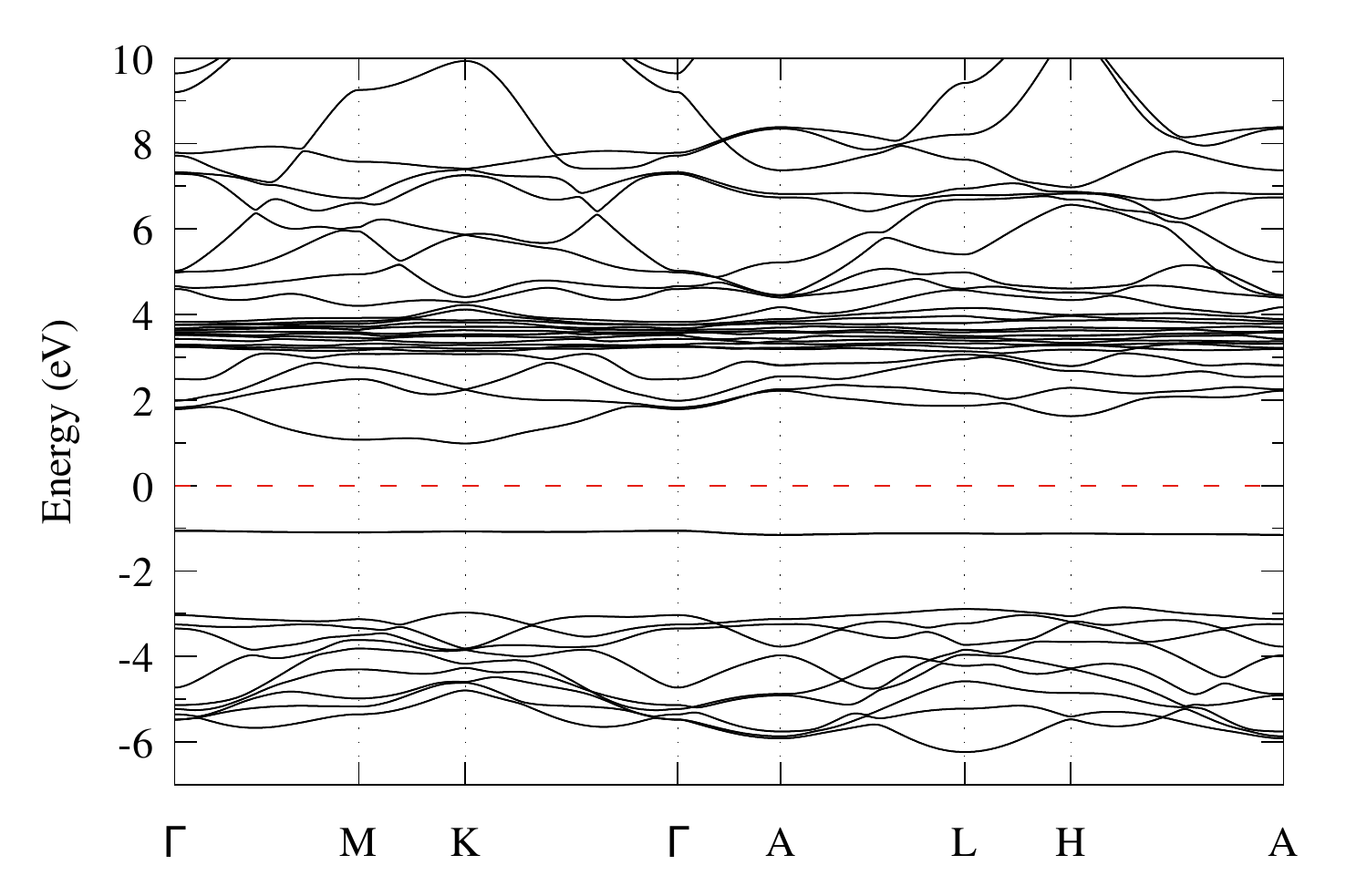}
\caption{Band structure of bulk Ce$_2$O$_3$ calculated using DFT+U. The Fermi level (dashed red line) is set to the origin of the energy axis.}
\label{Band_Ce2O3}
\end{figure}

In Figure \ref{fig:DOS2} we report the TDOS of bulk Ce$_2$O$_3$ obtained using an Hubbard correction in the range $U_\mathrm{eff}=[5:6]$ eV.
In Figs. \ref{Band_CeO2} and \ref{Band_Ce2O3} we plot the band structure of CeO$_2$ and Ce$_2$O$_3$, respectively, calculated using DFT+U.

\section{ELF in the optical limit: tests of different $U_\mathrm{eff}$ values. Momentum dispersion of the ELF}

\begin{figure*}[htbp!]
\centering
\includegraphics[width=0.7\textwidth]{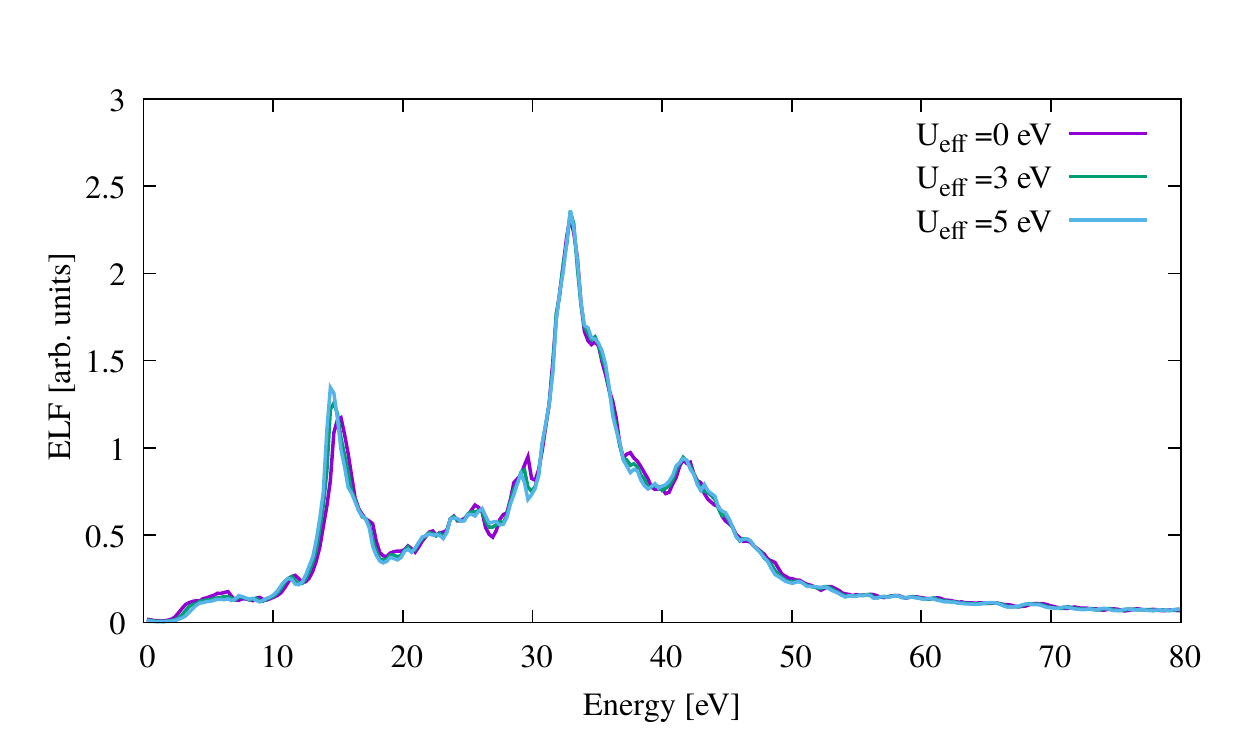}
\caption{ELF of bulk CeO$_2$ in the optical limit ($q\rightarrow 0$), calculated within the LSDA-ALDA approximation, using different values of the Hubbard correction $U_\mathrm{eff}$.}
\label{fig:EPSILON_B}
\end{figure*}
In Fig. \ref{fig:EPSILON_B} we report the ELF of bulk CeO$_2$ in the optical limit ($q\rightarrow 0$) analysing the impact of different values of the Hubbard correction $U_\mathrm{eff}$, concluding a negligible change. 

The impact of different approximations on the ELF to include many-body effects using either the random phase approximation (RPA) or the ALDA kernel, with and without Local Field Effects (LFE) is shown in Fig. 4 (top panel) of the main text. 

We notice that the key factor that mostly affects the accuracy is the introduction of LFE, since these effects are related to local density inhomogeneities and, thus, can be relevant in the assessment of the ELF as a function of the direction of the transferred momentum.

Furthermore, we also tested the bootstrap kernel \citep{Sharma2011}, that gives results similar to RPA showing that excitonic effects are small for this system. 

\begin{figure*}[htbp!]
\centering
\includegraphics[width=0.9\textwidth]{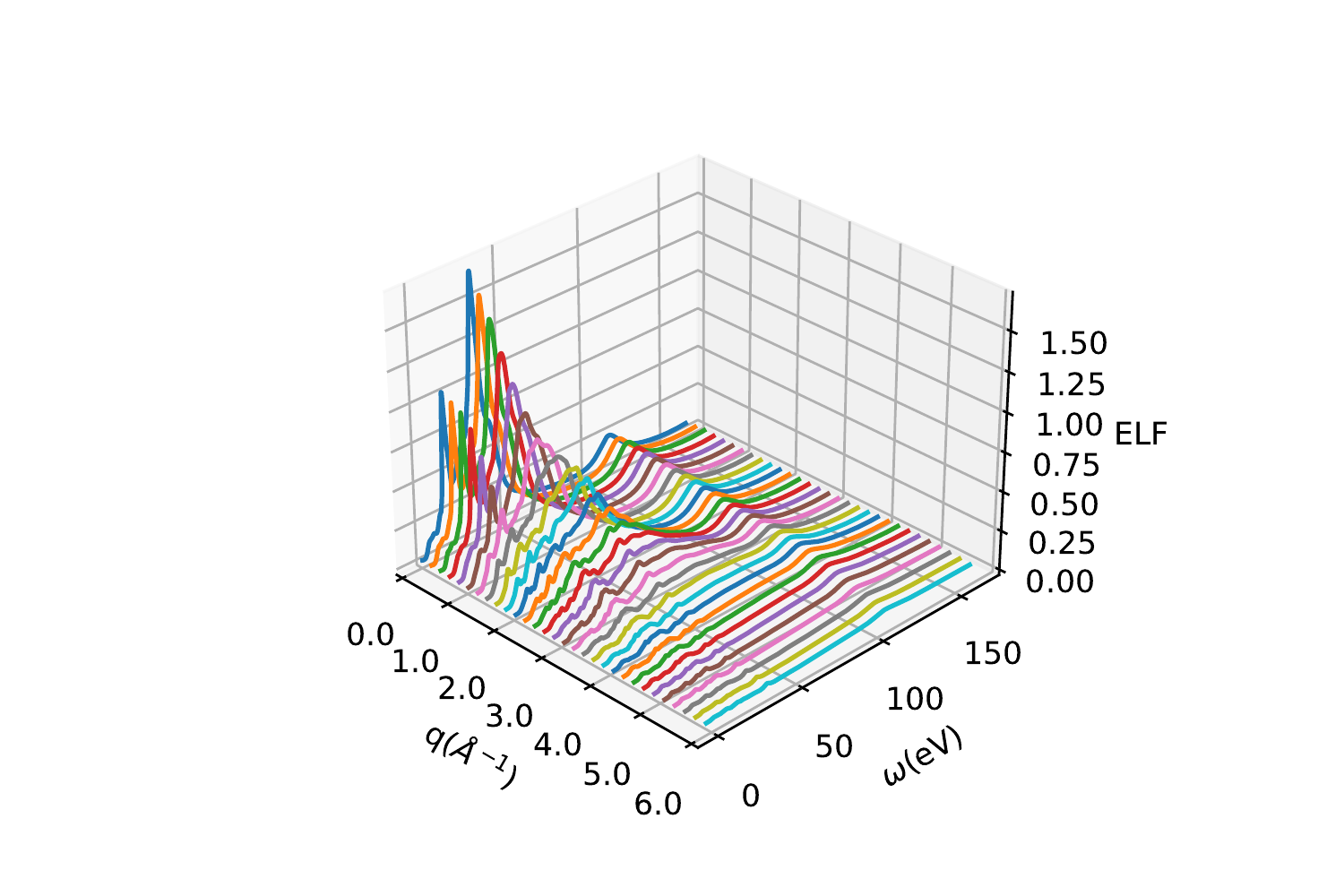}
\includegraphics[width=0.75\textwidth]{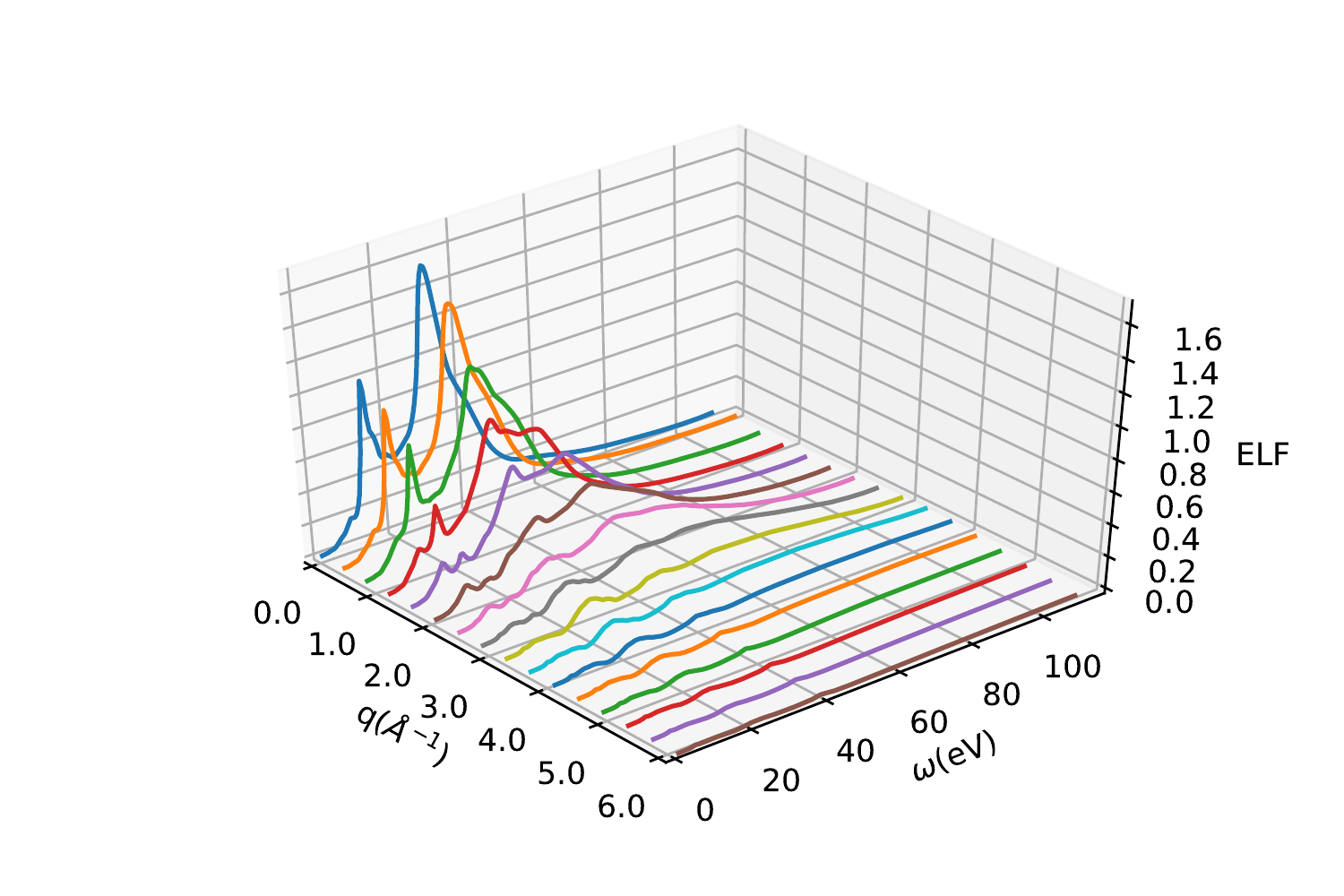}
\caption{ELF of bulk CeO$_2$ along the [111] direction (top panel) and of Ce$_2$O$_3$ along the [110] direction (bottom panel) from TDDFT calculations including LFE within ALDA.}
\label{fig:CeO2-highenergy}
\end{figure*}

The ab-initio Bethe surface of bulk CeO$_2$ up to $\sim$170 eV, including thus the $4d$ transitions, is presented in Fig. \ref{fig:CeO2-highenergy} (top panel).
It is interesting to notice that in the $100-120$ eV range the transition peak shifts towards lower energies with increasing momentum transfer. The Bethe surface of Ce$_2$O$_3$ is also shown in Fig. \ref{fig:CeO2-highenergy} (bottom panel).   

\begin{figure*}[htbp!]
\centering
\includegraphics[width=0.3\textwidth]{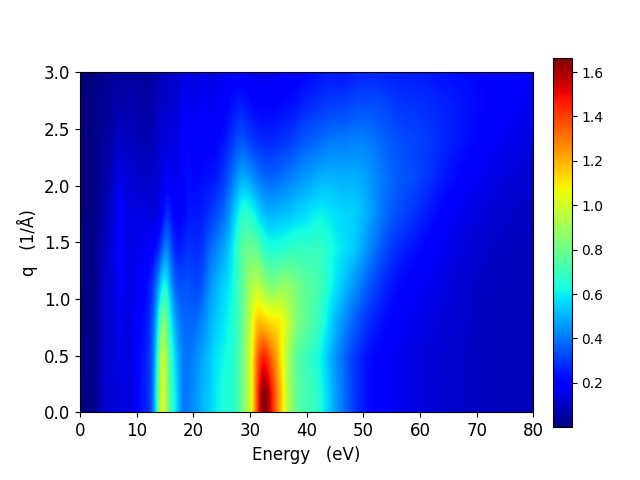}
\includegraphics[width=0.3\textwidth]{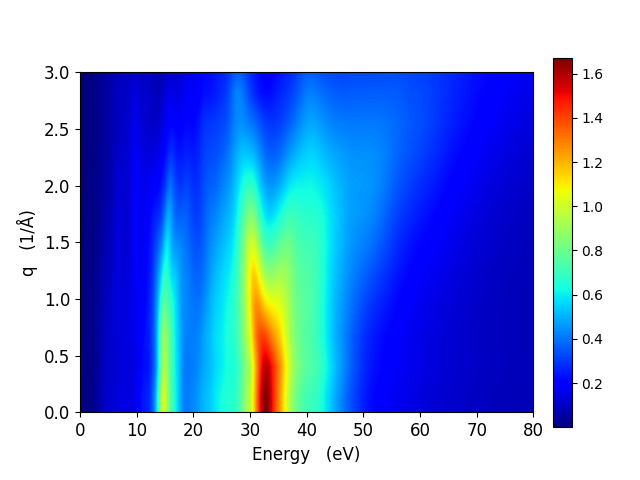}
\includegraphics[width=0.3\textwidth]{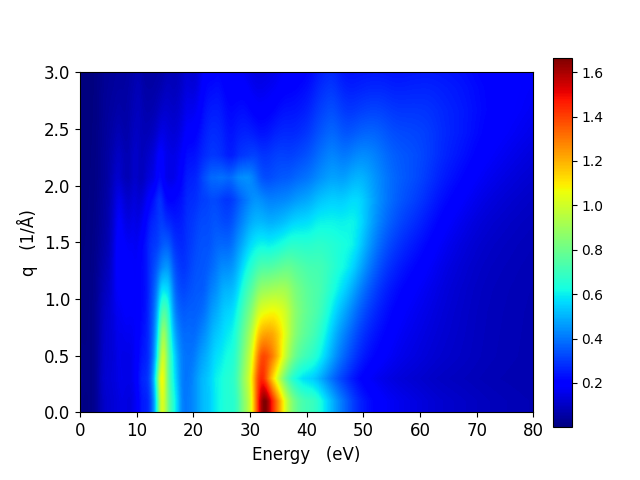}
\caption{Dependence of the ELF of bulk CeO$_2$ on different orientations of the momentum transfer vector. Left panel: [211]. Middle panel: [110]. Right panel: [111].}
\label{fig:CeO2dipq}
\end{figure*}

\begin{figure*}[htbp!]
\centering
\includegraphics[width=0.3\textwidth]{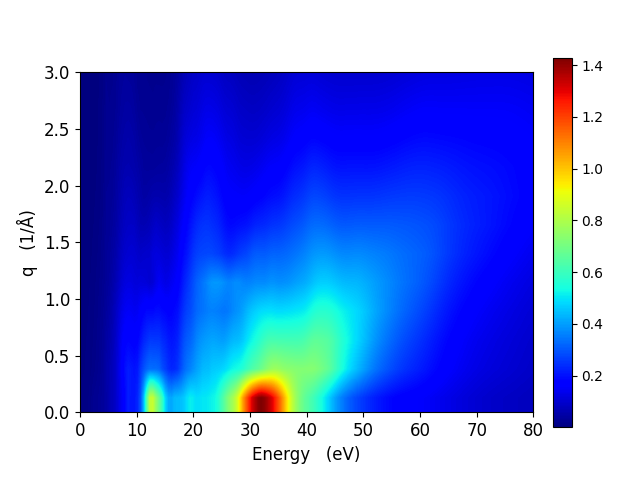}
\includegraphics[width=0.3\textwidth]{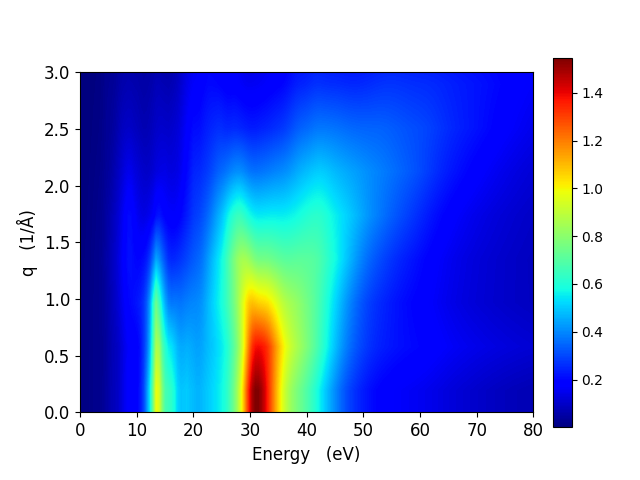}
\includegraphics[width=0.3\textwidth]{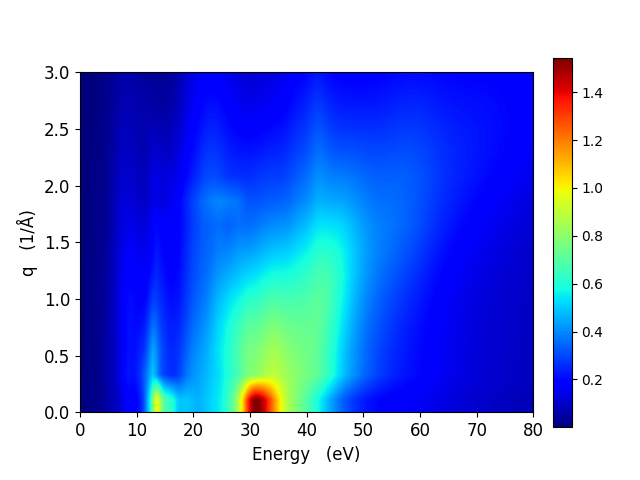}
\caption{Dependence of the ELF of bulk Ce$_2$O$_3$ on different orientations of the momentum transfer vector. Left panel: [001]. Middle panel: [110]. Right panel: [100].}
\label{fig:Ce2O3dipq}
\end{figure*}

\newpage
Dependence on the direction of the momentum transfer of the ELF of bulk CeO$_2$ and Ce$_2$O$_3$ is reported in Figs. \ref{fig:CeO2dipq} and \ref{fig:Ce2O3dipq}, respectively.




 
\section{Refractive index and extinction coefficient}

From the knowledge of the macroscopic dielectric function ($\overline{\epsilon}_1,\overline{\epsilon}_2$) of a material, one can reckon also its optical properties, such as the refraction index $n$ and the extinction coefficient $\kappa$, as follows:
\begin{equation}\label{nkappa}
n=\sqrt{\frac{1}{2}\Big(\sqrt{\bar{\epsilon}_1^2+\bar{\epsilon}_2^2} +\bar{\epsilon}_1\Big)},
\quad
\kappa=\sqrt{\frac{1}{2}\Big(\sqrt{\bar{\epsilon}_1^2+\bar{\epsilon}_2^2} -\bar{\epsilon}_1\Big),}
\end{equation}
where $\bar{\epsilon}_{1,2}$ are the real and imaginary part of the macroscopic dielectric function, respectively.
\begin{figure*}[htbp!]
\centering
\includegraphics[width=0.45\textwidth]{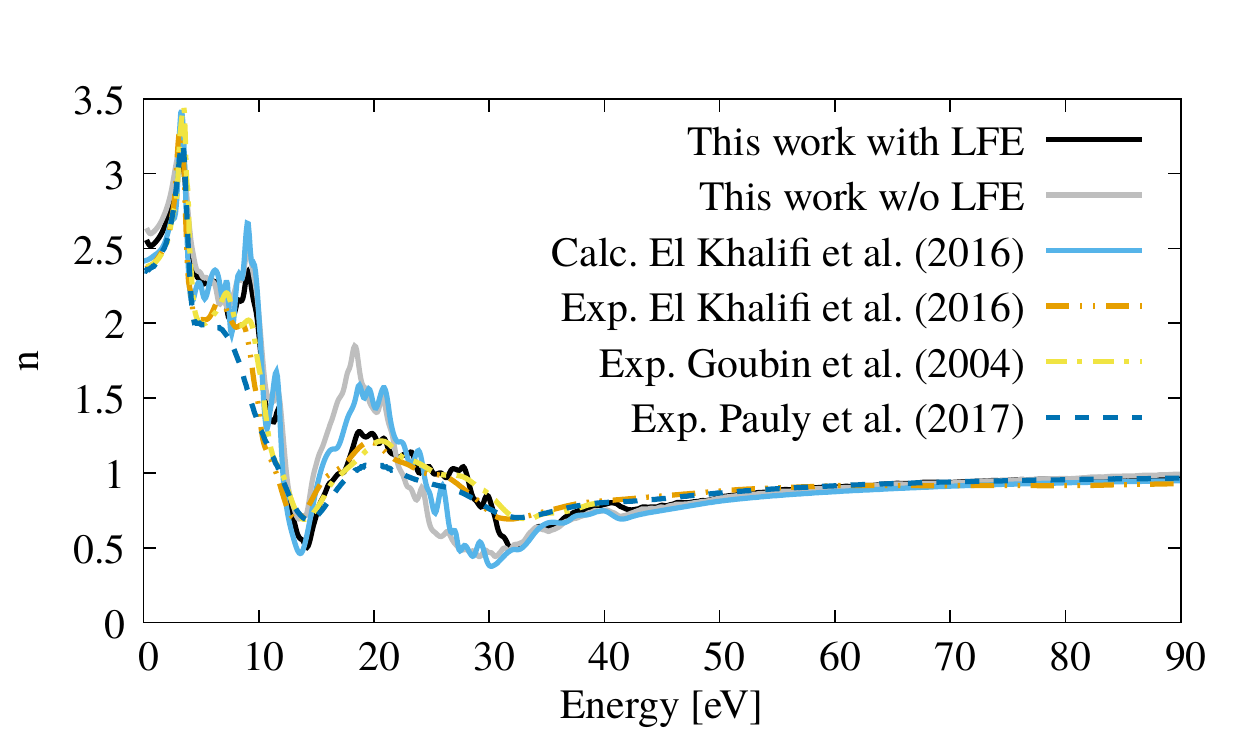}
\includegraphics[width=0.45\textwidth]{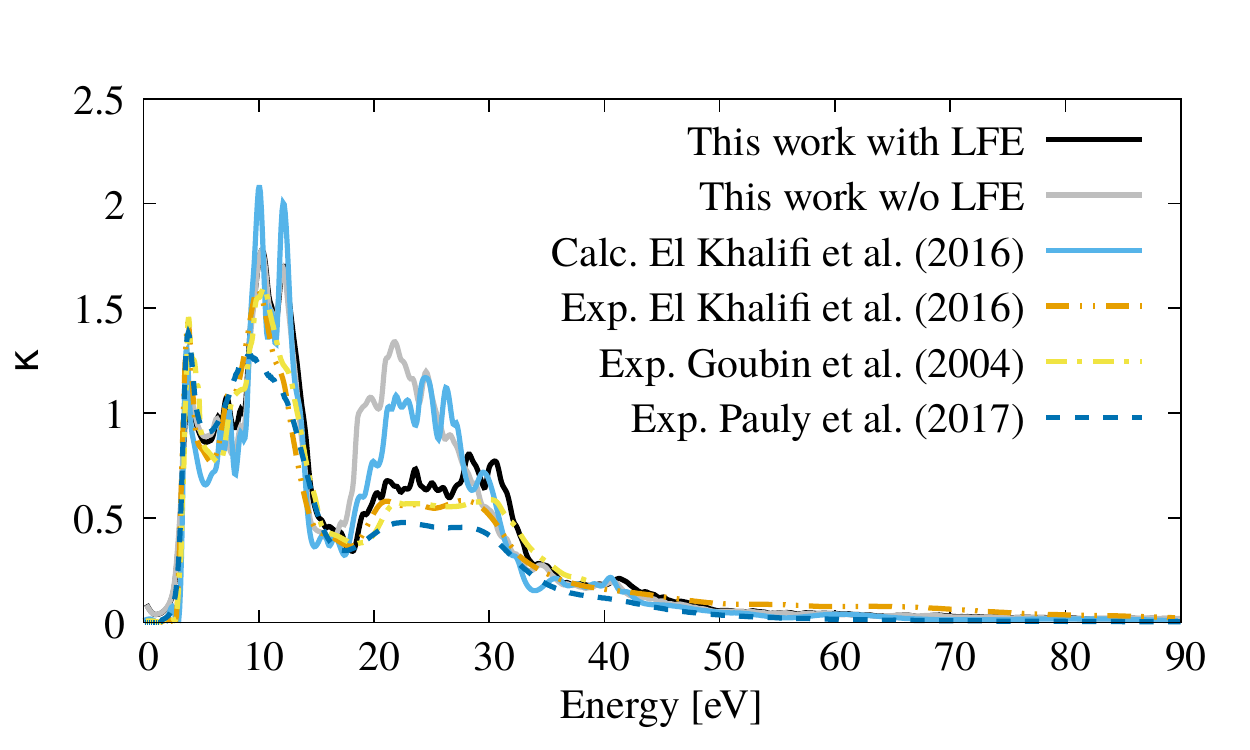}
\caption{Left: comparison between the refractive index of bulk CeO$_2$ calculated in this work and other experimental and computational studies \citep{Goubin2004,Khalifi2016,Pauly2017}. Right: comparison between the extinction coefficient of bulk CeO$_2$ calculated in this work and other experimental and computational studies \citep{Goubin2004,Khalifi2016,Pauly2017}.}
\label{fig:ComparisonRefractiveExp}
\end{figure*}

Using Eqns. (\ref{nkappa}), and the macroscopic dielectric matrix of eqn (4) 
in the main text 
we also calculated the refractive index and the extinction coefficient of both cerium oxides. In particular, we report in Fig. \ref{fig:ComparisonRefractiveExp} the refractive index (left panel) and the extinction coefficient (right panel) of bulk CeO$_2$, respectively, and their comparison with several other computational and experimental studies \citep{Khalifi2016, Goubin2004,Pauly2017}. We find an overall good agreement between our calculations and previous experimental and computational data when including LFE, with an appreciable difference emerging in the range $10-30$ eV between the calculated and experimental refractive indexes and in the range $20-30$ eV between the calculated and experimental extinction coefficients when switching-off LFE. In fact, LFE strongly suppress the intensity of the third major peak and lead to a better agreement of our simulations with the experimental data.

\begin{figure*}[htbp!]
\centering
\includegraphics[width=0.45\textwidth]{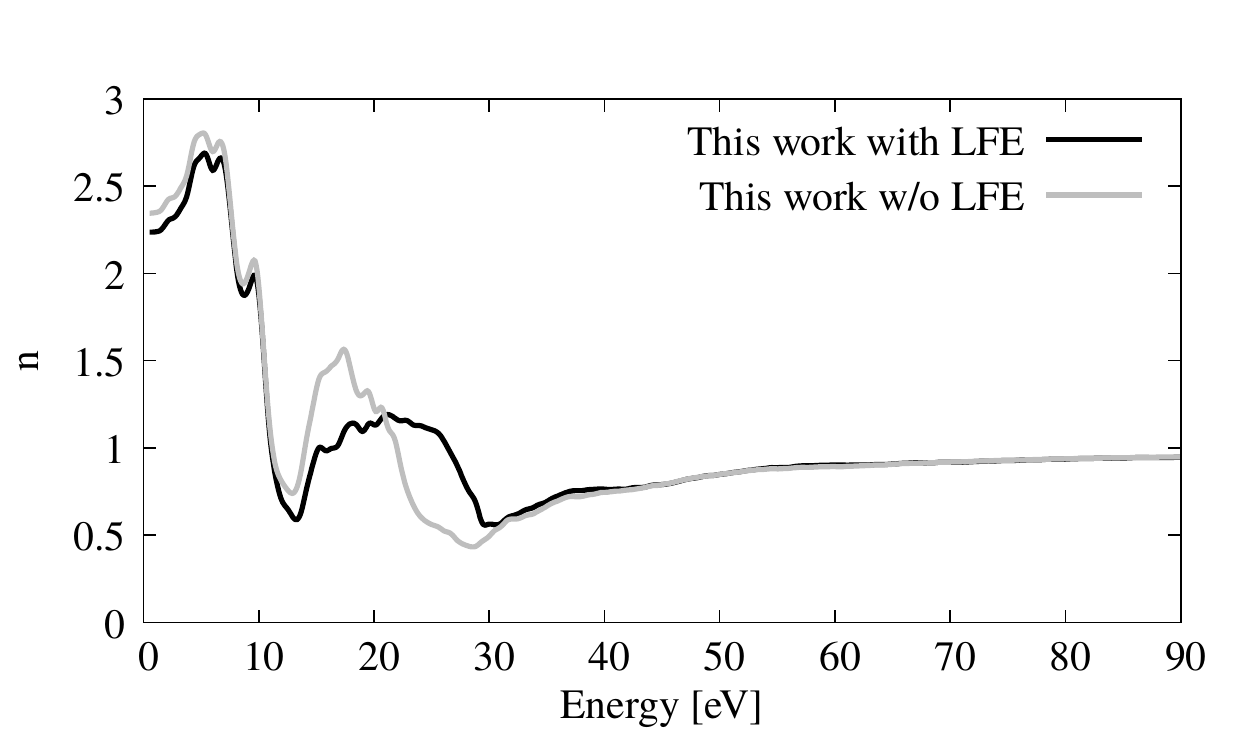}
\includegraphics[width=0.45\textwidth]{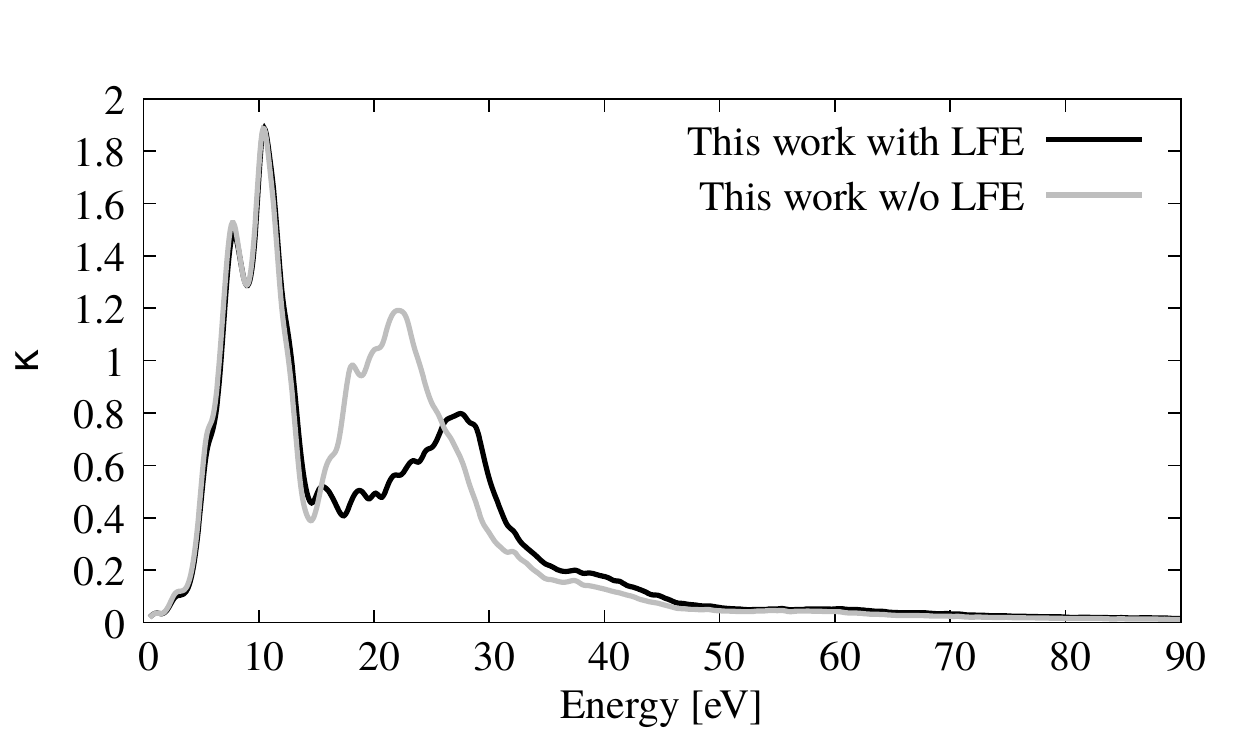}
\caption{Refractive index (left) and extinction coefficient (right) of bulk Ce$_2$O$_3$ with and without LFE.}
\label{fig:Refractive-Ce2O3}
\end{figure*}

Moreover, we report in Fig. \ref{fig:Refractive-Ce2O3} the refractive index (left panel) and the extinction coefficient (right panel) of bulk Ce$_2$O$_3$ with and without LFE.

\section{Parameters for the MELF-GOS fitting of the ab initio optical ELF}

Tables S1 and S2 report, respectively, the parameters used to fit the first principles optical ELF of CeO$_2$ and Ce$_2$O$_3$, by means of Eqns. (6)-(8) in the main text. In the column corresponding to $\Delta_i$, when the word ``Heaviside'' appears instead of a value, it means that the function $F$ is a Heaviside function for this function $i$. The criteria for the fitting, related to the fulfilment of the $f$-sum rules for individual transitions, are explained in the main text, sections 3.2 and 3.3.

\begin{table}[h]
\small
  \caption{\ Parameters used in the MELF-GOS fit of the optical ELF of bulk CeO$_2$ (see Eqns. (6-8) of the main text).}
\label{tbl:MELFGOS}
  \begin{tabular*}{0.8\textwidth}{@{\extracolsep{\fill}}lccccc}
 \hline
  Mermin    & $W_i$ (eV) & $\gamma_i$ (eV)   & $A_i$  (eV$^2$) &  $\Delta_i$ (eV$^{-1}$) & $W_{{\rm th},i}$ (eV) \\
    \hline\\
1   &   6.80 &  6.80 & 5.18 & Heaviside & 2.2                \\

2        &   15.10 & 3.54 & 44.13 &  1.873  & 5.2        \\
3 & 25.3 & 11.43 & 107.37 & 1.873  &  14.12  \\

4 & 33.20 & 7.62 & 125.88 & 1.873  & 14.12 \\
5 & 33.33 & 6.26 & 214.73 & 1.873  & 16.8 \\
6 & 42.18 & 14.97 & 259.16 & 1.873 & 37.80 \\
7 & 77.55 & 54.42 & 185.12 & Heaviside & 2.2 \\
8 & 113.34 & 14.97 & 303.59 & 0.184 & 100.68 \\
9 & 122.45 & 23.13 & 148.09 & 0.184 & 100.68 \\
10 & 326.54 & 408.17 & 296.18 & 0.184 & 206.53 \\
11 & 122.45 & 108.85 & 222.14 & Heaviside & 2.2\\
    \hline
  \end{tabular*}
\end{table}

\begin{table}[t]
\small
\caption{\ Parameters used in the MELF-GOS fit of the optical ELF of bulk Ce$_2$O$_3$ (see Eqns. (6-8) of the main text).}
\label{tbl:MELFGOS}
\begin{tabular*}{0.8\textwidth}{@{\extracolsep{\fill}}lccccc} \hline
Mermin  & $W_i$ (eV) & $\gamma_i$ (eV)   & $A_i$  (eV$^2$) &  $\Delta_i$ (eV$^{-1}$) & $W_{{\rm th},i}$ (eV) \\ \hline\\
1   &  8.44 &  2.72 & 1.63 & Heaviside & 2.04    \\
2   &  13.88 & 2.18 & 19.25 &  Heaviside & 3.83  \\
3 & 16.6 & 7.62 & 51.83 & Heaviside &  3.83  \\
4 & 31.97 & 10.34 & 251.76 & 0.735 & 13.06 \\
5 & 31.97 & 8.16 & 192.52 & 0.735 & 15.10 \\
6 & 38.10 & 14.69 & 255.46 & 0.735 & 37.80 \\
7 & 64.49 & 32.65 & 162.90 & Heaviside & 2.04 \\
8 & 129.93 & 15.24 & 266.57 & 0.184 & 100.68 \\
9 & 137.15 & 10.07 & 133.28 & 0.184 & 100.68 \\
10 & 326.54 & 408.17 & 296.18 & 0.184 & 206.53 \\ \hline
\end{tabular*}
\end{table}

\end{document}